\documentclass[aps,12pt,nofootinbib,preprint]{revtex4}
\pdfoutput=1
\usepackage{pdfsync}
\usepackage{textcomp}
\usepackage{amssymb}
\usepackage{anysize}
\usepackage{bold-extra}
\usepackage{enumerate}
\usepackage{graphicx}
\usepackage{lscape}
\usepackage{mathrsfs}
\usepackage{multirow}
\usepackage[final]{pdfpages}
\usepackage{rotating}
\usepackage{subfig}
\usepackage{url}
\usepackage{wrapfig}
\usepackage{amsmath}
\usepackage{fancyhdr}
\usepackage{verbatim} 
\usepackage[linktocpage,colorlinks]{hyperref}
\usepackage{color}
\def\nn{\nonumber}
\def\bea{\begin{eqnarray}}
\def\eea{\end{eqnarray}}
\def\ba{\begin{eqnarray}}
\def\ea{\end{eqnarray}}
\def\be{\begin{equation}}
\def\ee{\end{equation}}
\def\beq{\begin{equation}}
\def\eeq{\end{equation}}

\def\lsim{\mbox{\raisebox{-.6ex}{~$\stackrel{<}{\sim}$~}}}

\newcommand\hc{\text{h.c.}}
\def\({\left(}
\def\){\right)}
\def\VEV#1{\langle #1\rangle}

\def\CAf{{\mathcal{A}_{\text{fwd}}}}
\def\CA{{\mathcal{A}}}
\definecolor{Ben}{cmyk}{0.2,0.4,0.7,0.}

\definecolor{MyLightMagenta}{cmyk}{0.1,0.8,0,0.1}
\begin{document}

\preprint{UCSD/PTH 13-15}
\title{Theoretical Constraints on Additional Higgs Bosons in Light of the 126 GeV Higgs}
\author{Benjam\'{i}n Grinstein}
\email{bgrinstein@ucsd.edu}
\affiliation{Department of Physics, University of California, San Diego, La Jolla, CA 92093 USA}
\author{Christopher W. Murphy}
\email{cmurphy@physics.ucsd.edu}
\affiliation{Department of Physics, University of California, San Diego, La Jolla, CA 92093 USA}
\author{David Pirtskhalava}
\email{david.pirtskhalava@sns.it}
\affiliation{Scuola Normale Superiore, Piazza dei Cavalieri 7, 56126 Pisa, Italy}
\author{Patipan Uttayarat}
\email{uttayapn@ucmail.uc.edu}
\affiliation{Department of Physics, University of Cincinnati, Cincinnati, OH 45220 USA}
\affiliation{Department of Physics, Srinakharinwirot University, Wattana, Bangkok 10110 Thailand}

\begin{abstract}
We present a sum rule for Higgs fields in general representations under $SU(2)_L \times U(1)_Y$ that follows from the connection between the Higgs couplings and the mechanism that gives the electroweak bosons their masses, and at the same time restricts these couplings.
Sum rules that follow from perturbative unitarity will require us to include singly and doubly charged Higgses in our analysis. We examine the consequences of these sum rules for Higgs phenomenology in both model independent and model specific ways. The relation between our sum rules and other works, based on dispersion relations, is also clarified.
\end{abstract}

\maketitle

\section{Introduction}
The properties of the narrow resonance that has been discovered at the
LHC is fit well by the Standard Model's (SM's) Higgs particle
hypothesis. With the measured mass as input, the production rate and
decay branching fractions are precisely predicted in the SM and
compare well with the experiments. 

In this note we ask a simple question: is it theoretically possible 
to have additional Higgs-like particles? We are interested in 
additional scalars that may have a mass different from 126~GeV but
with similar production cross section and similar decay width into
$\gamma\gamma$ and $WW$ final states. Alternatively one can ask, if
such a particle (or particles) exist, what are the constraints on
their properties? Are there model independent constraints?

The question may be more than academic. CMS has made public a
note~\cite{CMS-PAS-HIG-13-016} that points to a hint of a resonance at
about 136~GeV observed in the $\gamma\gamma$ channel, which is produced both by gluon fusion (ggF) and vector boson fusion (VBF), with
signal strength close to unity for both. Below we refer to the
additional resonance as the~$h'$.\footnote{In addition, another CMS
  analysis~\cite{CMS-PAS-HIG-13-007}  has a slight excess in the
  dielectron channel around 134 GeV. The observed 95\%
  confidence-level upper limit on the cross section times branching
  ratio is 0.048 pb at 134 GeV.  Assuming the production rate to be the
  same as for the SM Higgs, this corresponds to an upper limit on the
  branching ratio of $\text{Br}(h^{\prime} \to e^- e^+) = 0.0025$. The
  background-only expected limit is 0.0015.} Moreover, ATLAS and CMS
claim to rule out heavier Higgs bosons through searches in the $WW$
and $ZZ$
channels~\cite{ATLAS-CONF-2013-013,ATLAS-CONF-2013-067,CMS-PAS-HIG-12-024,CMS-PAS-HIG-13-002,CMS-PAS-HIG-13-003,CMS-PAS-HIG-13-008,CMS-PAS-HIG-13-014}. Naively
combining these searches rules a 
neutral Higgs particle with  mass in the range 128~GeV to 1000~GeV and
SM interaction strength; see Appendix~\ref{sec:app}. It is important to understand the generality
of the assumptions for this limit, which we discuss in what follows.

Consider in some more detail the reported excess in the diphoton
channel, seen  in both ggF and VBF production modes.  Following the procedure outlined in
Ref~\cite{Azatov:2012bz,Azatov:2012qz} we
extract the production cross-section times branching ratio from the
CMS exclusion limit:
\begin{equation}
\begin{aligned}
	\sigma_{ggF}\times BR(h^\prime \rightarrow \gamma\gamma) & = 0.036 \pm 0.013 \text{ pb},\\
	\sigma_{VBF+Vh^\prime}\times BR(h^\prime \rightarrow
        \gamma\gamma) & =  0.007 \pm 0.003 \text{ pb}.
\end{aligned}
\end{equation} 
Dividing these cross sections by the SM prediction for a Higgs boson with $m_{h^{\prime}} = 136.5$ GeV yields signal strengths, $\mu \equiv (\sigma \times \text{Br}) / (\sigma_{SM} \times \text{Br}_{SM})$, of $\mu_{ggF} = 1.1 \pm 0.4$ and $\mu_{VBF} = 1.6 \pm 0.7$ respectively. Let's characterize the coupling of $h^\prime$ to vector
bosons and fermions by
\begin{equation}
\label{eq:lag-prime}
  \mathcal{L}_{h^\prime} \supset  \frac{2m_W^2}{v_{\text{EW}}}a_{h^\prime} W^\mu W_\mu h^\prime   +\frac{m_Z^2}{v_{\text{EW}}}a_{h^\prime} Z^\mu Z_\mu h^\prime - \sum_i \frac{m_{f_i}}{v_{\text{EW}}} c_{f_i}^\prime \bar{f}_i f_i h^\prime
\end{equation}
where $v_{\text{EW}}=(\sqrt2G_F)^{-1/2}\approx 246~\text{GeV}$ is the electroweak VEV. We assume (approximate) custodial symmetry, hence the same couplings $a_{h'}$ to $W$ and $Z$. 
See Sec. 3A for a discussion of the constraints electroweak precision data puts on custodial-violating theories. 

\begin{figure}
  \centering
 \includegraphics[width=0.5\textwidth]{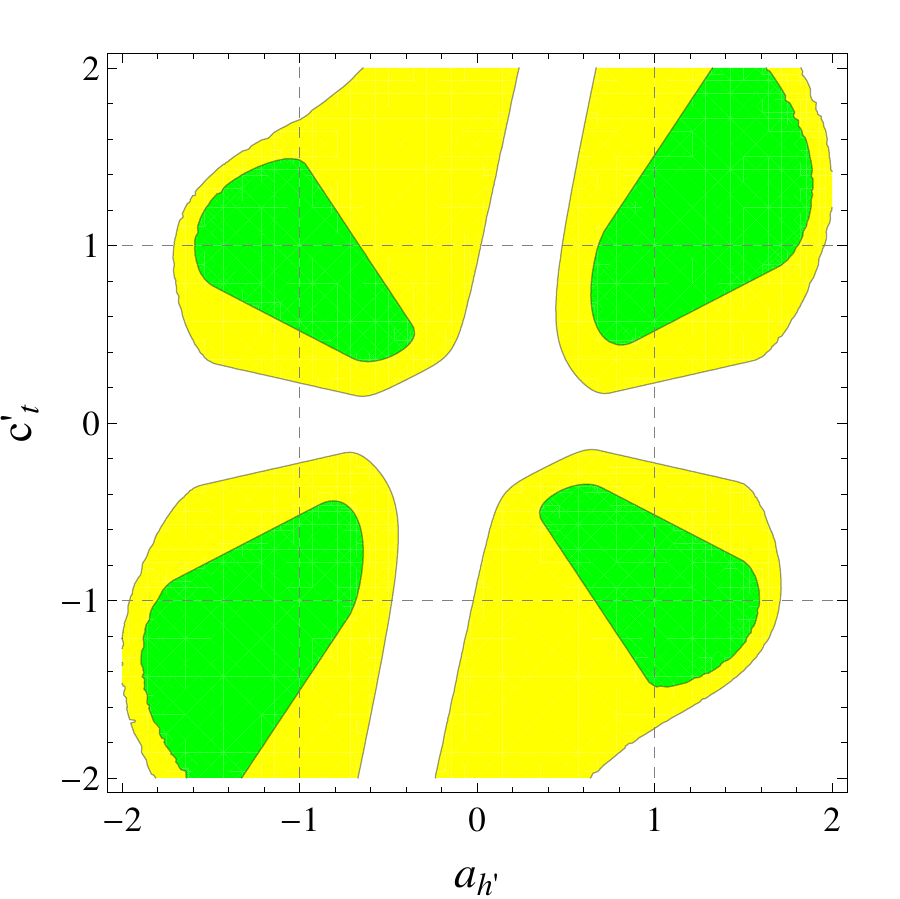}
\caption{Model independent analysis of the $h^\prime$ couplings.  The
  green and yellow regions are compatible with the CMS hint of a
  136~GeV Higgs resonance  at 68\% and 95\% CL, respectively. }
 \label{fig:136}
\end{figure}

The parameters in Eq.~\eqref{eq:lag-prime} can be estimated from these
two measurements by performing a $\chi^2$ fit to the
data. Since there are more parameters than measurements, there
should be at least one set of parameters that exactly reproduces the
measurements.  We  assume that $h^{\prime}$ can only decay to SM particles, and
that $\left|c_{f \neq t}^{\prime}\right| \leq 3$. The couplings we are 
most interested in are $a_{h^{\prime}}$ and $c_t^{\prime}$, so we project the allowed parameter
space onto the $a_{h^{\prime}} - c_t^{\prime}$ plane. Note that we are not performing a goodness-of-fit test, but
are simply trying to estimate parameters. The result is shown in
Fig.~\ref{fig:136}. The green and yellow regions correspond to
$\chi^2 \le 2.30$ and $6.18$, hence are compatible with
the CMS measurements at the 68\%  and 95\% confidence levels (CL),
respectively.
 
\begin{figure}
  \centering
  \subfloat[]{\label{fig:136cwma}\includegraphics[width=0.48\textwidth]{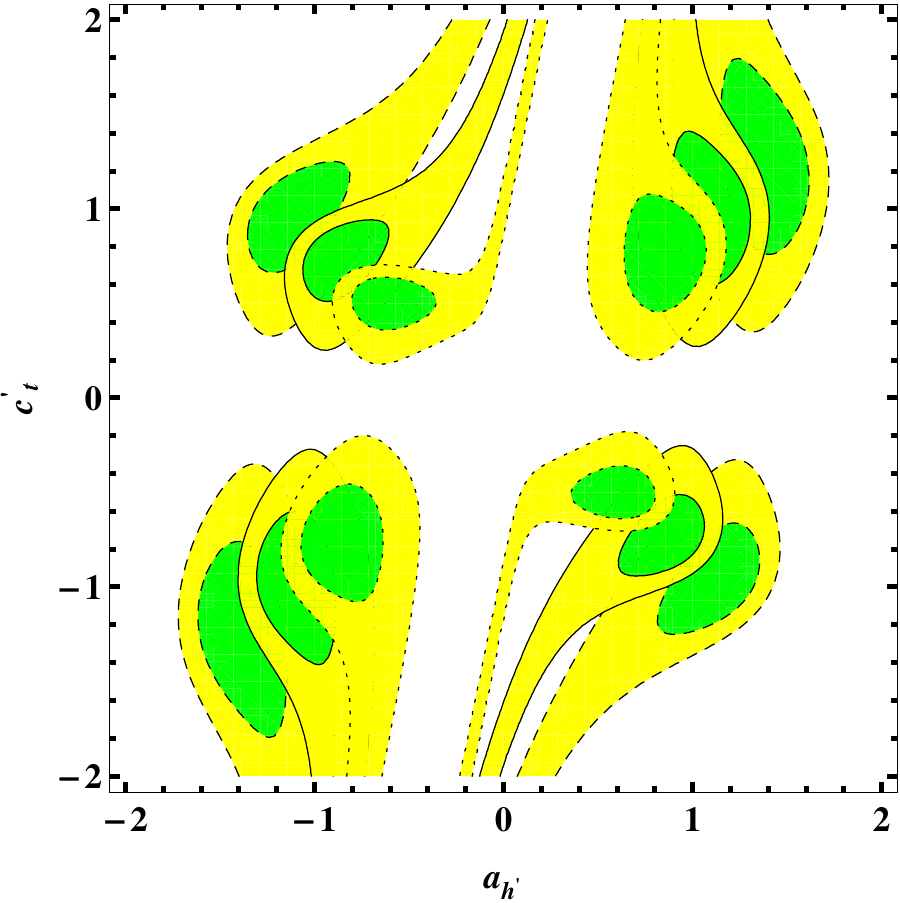}} \hspace{5pt}
  \subfloat[]{\label{fig:136cwmb}\includegraphics[width=0.48\textwidth]{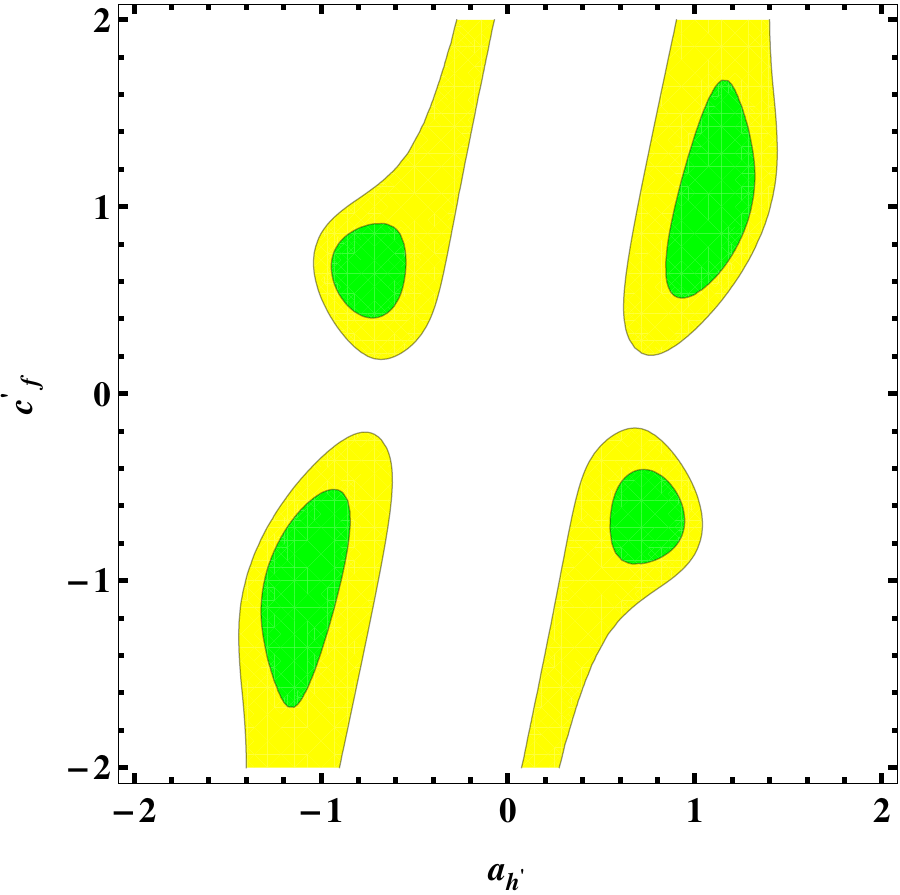}}
  \caption{The preferred values of $(a_{h^{\prime}}, c^{\prime}_t)$
    for interpreting the bump in the CMS data as a second Higgs
    boson. The regions of parameter space that are allowed at 68\% and 95\%~CL 
are shown in green and yellow, respectively. In
    Fig.~\ref{fig:136cwma}, the couplings of $h^{\prime}$ to fermions
    other than the top-quark are fixed to a common value. The dotted,
    solid, and dashed contours correspond to $c^{\prime}_{f \neq t} =
    \{0, 1, 2\}$ respectively. In Fig.~\ref{fig:136cwmb}, the
    signal strength modifier for fermions is assumed to be
    universal. }
  \label{fig:136cwm}
\end{figure}

Alternatively, it is instructive to see what happens when an ansatz is
made for the other parameters in the model.  In
Fig.~\ref{fig:136cwma}, the couplings of $h^{\prime}$ to fermions
other than the top-quark are fixed to a common value. The dotted,
solid, and dashed contours correspond to $c^{\prime}_{f \neq t} = \{0,
1, 2\}$ respectively, while in Fig.~\ref{fig:136cwmb}, the signal
strength modifier for fermions is assumed to be universal. The latter
is an interesting case because there is a class of models, which we
discuss in depth below, where there is a single common $c_f^{\prime}$
for all fermions.

We see that CMS measurements prefer sizable coupling of $h^\prime$ to
both the top and the vector-bosons.  The question we posed above can be
stated more specifically: how much of the allowed region in
Fig.~\ref{fig:136} is compatible with the established properties of
the 126 GeV Higgs resonance?  We hasten to indicate that we address this
question in generality, not just as it may pertain to a putative state
at 136~GeV (but we do use the 136~GeV CMS data as an instructive example).

What precisely do we mean by ``Higgs-like'' particles? What makes a Higgs-like
particle, or Higgs for short, special is its tri-linear coupling to
electroweak vector bosons, say $hW^+W^-$ or $hZZ$. Indeed, in a gauge
theory all fields, $\psi$, except the Higgs,\footnote{And other
  electroweak vector bosons, of course.} have couplings to gauge bosons,
$A$, with the field appearing quadratically, $\psi^2A$ or
$\psi^2A^2$. Hence a unique characteristic of Higgs particles is that
they can be produced in $s$-channel vector boson fusion, and can decay
into pairs of vector bosons. This generalized definition of Higgs
particle includes, of course, singly and doubly charged particles in
addition to the  more familiar neutral (CP-even) Higgs. In fact,
considerations of perturbative unitarity will require that we include
singly and doubly charged Higgs in our analysis. 

We will derive a number of sum rules that will restrict the couplings
of the Higgs particles. The sum rules are model independent, but
derived only at tree level. For each sum rule we derive we will show
in explicit examples how they are saturated. We will see, model independently,
that not all the allowed region in Fig.~\ref{fig:136} is compatible
with the established properties of the 126 GeV Higgs resonance.  We
will also see that the region in Fig.~\ref{fig:136} compatible with
several specific models of the $h'$ is further restricted.  

Sum rules for Higgs
particles have been considered before. There is a vast literature on 2HDM models, recasting explicit results as sum rules, see e.g. \cite{Grzadkowski:1999wj,Ginzburg:2004vp,Celis:2013rcs,Celis:2013ixa}
for an incomplete list.
Sum rules for the couplings of arbitrary number of Higgses in general representations of the electroweak group were first derived by Gunion, Haber, and Wudka using perturbative unitarity arguments in \cite{Gunion:1990kf}. To make the presentation self-contained, we review below the derivation of these sum rules. In addition, we present bounds on combinations of masses and couplings of the Higgs particles, that to the best of our knowledge have not been considered for general Higgs sectors before\footnote{While sum rules for couplings are satisfied automatically for a renormalizable theory with an arbitrary Higgs sector, the sum rules bounding masses of Higgs particles carry extra non-trivial information implied by perturbative unitarity.}. These perturbative unitarity mass bounds are the multi-Hiiggs generalization of the celebrated result by Lee, Quigg and Thacker \cite{Lee:1977eg} that placed an upper bound on the higgs mass of $\sim$700 GeV. In Ref.~\cite{Distler:2006if} a
twice subtracted dispersion relation for longitudinal $WW$ scattering
was obtained and applied to Higgsless models. In
Ref.~\cite{Low:2009di,Falkowski:2012vh,Urbano:2013aoa}, a similar relation was given for a model
with a single Higgs particle with non-standard couplings. In particular, it was shown that the couplings of the light Higgs to a $W$ pair obeys the following simple dispersion relation
\beq
\label{eq:disp}
1-a^2_h=\frac{v^2}{\pi}\int^\infty_0 \frac{ds}{s}
\(\sigma_{+-}(s)-\sigma_{++}(s)\)~, \eeq where $\sigma_{+-}$ denotes
the cross section for a longitudinal $W$ pair annihilation,
$W^{L+}W^{L-}\to \textit{anything}$, and similarly for
$\sigma_{++}$. Moreover, it was noticed in
Ref.~\cite{Low:2009di,Falkowski:2012vh} that the last relation implies
that enhanced Higgs~$\to WW$ couplings require doubly
charged states (that couple to vector bosons) be present in the
theory.

We will generalize \eqref{eq:disp} to a multi-Higgs case below. This
dispersion relation holds under the assumption of unitarity of the
full UV theory (supplemented by a more technical assumption that the
Froissart bound is sufficiently unsaturated). While \eqref{eq:disp} is
true to all orders in the loop expansion and for nonperturbative
theories as well as perturbative ones, in what follows we will be
exclusively interested in the tree-level amplitudes in perturbative
theories with definite UV field content. Our tree-level sum rules will
then guarantee that the assumption of (perturbative) unitarity under
which \eqref{eq:disp} holds is satisfied at order $\hbar^0$.  We
discuss how exactly our sum rules are consistent with \eqref{eq:disp}
in detail in Sec. \ref{disper-rels} and App. \ref{sec:appb}.

The paper is organized as follows. In Sec.~\ref{multidoublet} we
present a simple sum rule for multi-Higgs doublet models that follows
from the connection between the Higgs couplings and the mechanism that
gives the electroweak bosons their masses, and we generalize this to
models with Higgs fields in other representations of $SU(2)_L \times U(1)_Y$ 
in Sec.~\ref{gener}. These sum rules are model dependent, so in
Sec.~\ref{sec:unitarity} we turn to sum rules that follow from
perturbative unitarity. We examine the consequences of these sum rules
on the allowed region in Fig.~\ref{fig:136} in
Sec.~\ref{model-indep}. We then study in some detail the phenomenology
of some specific models in Sec.~\ref{pheno}.  Finally we study the
relation between our sum rules and other work, based on dispersion
relations in Sec.~\ref{disper-rels} and offer some concluding remarks
in Sec.~\ref{concl}. 
To make the paper easily accessible, we list the physical Higgses couplings in App.~\ref{sec:app0}. 
We collect the Higgs data used in our analysis in App.~\ref{sec:app}. 
App.~\ref{sec:appb} is devoted to detailed derivation of the dispersion relation.  

\section{multi-Higgs Doublet model}
\label{multidoublet}
The CMS note points out that the $h'$ is incompatible with a two Higgs
doublet model (2HDM) hypothesis (see, e.g., \cite{Grinstein:2013npa,
  Barbieri:2013hxa,Barbieri:2013nka} and references therein for a
recent general analysis of Type-II 2HDM and (N)MSSM Higgs
sectors). The reason is this. Since both the 126~GeV and 136~GeV
states couple with similar strength to $WW$, we can immediately
discount the CP-odd neutral Higgs as one of these states. Assuming the
light CP-even neutral Higgs is the particle observed at 126~GeV, fits
of Higgs data to the 2HDM assumption give $\alpha+\beta\approx \pm
\pi/2$, where $\alpha$ is the mixing angle between CP-even neutral
Higgs mass eigenstates and $\tan\beta= v_1/v_2$ is the ratio of the
vacuum expectation values (VEV) of the two doublets. The ratio of the
coupling of the heavier Higgs to $WW$ to that of the lighter Higgs is
$\cot(\alpha+\beta)$, so the fit gives a very suppressed $h'$ to
$W^+W^-$ coupling.

This observation is readily generalized to models with any number of
Higgs doublets. Consider a model with $N$ Higgs doublets, $H_i$, all
with $Y=\tfrac12$, with VEV $v_i/\sqrt2$ and CP even neutral scalar
$\tilde{h}_i$. The kinetic energy terms of these doublets give the $W$-mass and
the couplings of the $\tilde{h}_i$ to $W^+W^-$:
\beq
\label{WWhdoublets}
\mathcal{L}=\cdots+\tfrac14g^2W^+_\mu W^-_\nu\eta^{\mu\nu}\sum_{i=1}^N(v_i+\tilde{h}_i)^2.
\eeq
This can be interpreted as follows. The vector
$\tfrac12g^2(v_1,\ldots,v_N)\equiv  gM_W\vec {\tilde{ a}}=g M_W ( \tilde{a}_1,
\ldots, \tilde{a}_N)$ characterizes the couplings of the fields $\vec{\tilde{
h}}=(\tilde{h}_1,\ldots,\tilde{h}_N)$ to $W^+W^-$, and the norm of the vector is fixed,
$|\vec{\tilde{a}}|^2=1$. The fields in $\vec{\tilde{h}}$ do not in general
correspond to mass eigenstates. An orthogonal rotation $\vec{\tilde{h}}=
R\vec{h}$ brings the mass matrix to diagonal form. The mass
eigenstates couple to $WW$ with strength $\vec{ a}=R^T\vec{
\tilde{a}}$. Since $R$ is orthogonal $|\vec{\tilde{a}}|^2=1$ implies $|\vec
{a}|^2=1$. Without loss of generality we take $
h_1$ to correspond to the 126~GeV observed resonance. Then
\beq
\label{sumrule1}
\sum_{i>1} a_i^2=1- a_1^2\,.
\eeq
That is, the CP-even neutral Higgs resonances
other than $ h_1$ (the observed 126~GeV one) can couple to $WW$
only to the extent that the coupling of $ h_1$ to $WW$ differs
form that of Higgs in the (one Higgs doublet) SM. 

\section{Generalizations}
\label{gener}
The results of the previous section can be generalized to the case of
Higgs fields in any non-trivial representation of $SU(2)$. This may
seem as only of academic interest since electroweak precision data
(EWPD) places stringent constraints on the VEV of non-doublet
representations. But there are exceptions, like the model of Georgi
and Machacek \cite{Georgi:1985nv} and models with \mbox{isospin-3}
fields \cite{Kanemura:2013mc}. We carry out
the analysis in general, and return below to considerations of EWPD.

In the case of arbitrary representations, Eq. \eqref{WWhdoublets} is
replaced by
\beq
\label{eq:genrlzdWWh}
\mathcal{L}=\cdots+\tfrac14g^2W^+_\mu W^-_\nu\eta^{\mu\nu}\sum_{i=1}^NC_i(v_i+\tilde{h}_i)^2,
\eeq
where we assume $\tilde{h}_i$ is the CP-even neutral scalar from the
$(2n_i+1)$-dimensional representation, with the third component of isospin given by $m_i$, so
that 
\beq 
\label{eq:cgiven}
C_i=C(n_i,m_i)=2[n_i(n_i+1)-m_i^2].  
\eeq
The couplings of $\tilde{h}_i$ are now characterized by $\vec{\tilde{a}}=\tfrac{g}{2M_W}( C_1
v_1,\ldots,C_N v_N)$ and those of mass eigenstates by $\vec{
  a}=R^T\vec {\tilde{a}}$, with $R^TR=1$. The constraint 
\beq
\label{sumrule2}
\sum_{i=1}^N \frac{\tilde{a}_i^2}{C_i} = 1,
\eeq
is an ellipse in $\vec{\tilde{a}}$ space. Rotating to $\vec{a}$ space,
this remains an ellipse. Now suppose one of the couplings in
$\vec{ a}$ has been measured. Without loss of generality we take
this to be the first component, $ a_1$. For $a_1$ sufficiently close to $1$, the least constraining
case is when the rotation $R$ makes $ a_1$ line up with the
semi-major axis. Consider, for example,  the $N=2$ case with $C_1=1$ and
$C_2>1$. Then we must have 
\beq
\label{eq:exteremexample}
 a_2^2= 1 - \frac{ a_1^2}{C_2}~.
\eeq
More generally, the constraint on the coupling of the mass eigenstates
to $WW$ takes the form
\beq
\label{eq:sum0}
a^T R^T C^{-1}R a = 1,
\eeq
where $C$ is the diagonal matrix of the $C_i$ and we have used vector
notation (with $a=\vec{a}$). This in principle alleviates constraints on the magnitude of couplings of extra neutral Higgs bosons to $W$ and $Z$.

While the rotation to the mass eigenstate may greatly relax the
constraint on the coupling of the second Higgs, the coupling of the
first Higgs resonance to fermions is correspondingly
reduced. Only \mbox{isospin-$\tfrac12$} states can couple directly to fermions. If only
$H_1$ is in the doublet representation then the coupling to fermions
is through
\beq
\mathcal{L}_f=  \tilde H_1 \bar q_L \lambda_U u_R + H_1 \bar q_L
\lambda_D d_R+H_1 \bar \ell_L \lambda_E e_R ,
\eeq
where $\tilde H_1=i\sigma^2 H_1^*$ and $\lambda_{U,D,E}$ are the
matrices of Yukawa couplings. Expanding this about the vacuum and
retaining only the couplings of the neutral Higgs mass eigenstates one
obtains 
\beq
\mathcal{L}_f=\frac1{v_1}R_{1i} h_i\left(\bar u m_U u +
  \bar d m_D d + \bar e m_E e\right)~,
\eeq
where $m_{U,D,E}$ are the mass matrices.  Orthogonality of $R$ implies
$\sum_{i>1} R^2_{1i}=1-R^2_{11}$, limiting the extent to which the
remaining Higgs resonances couple to fermions. Returning to the simple
example above, in the extreme case that the doublet field $C_1=1$ is
maximally rotated with a non-doublet, $C_2>1$ as given by
Eq.~\eqref{eq:exteremexample}, one has $R_{11}=0$ and $R_{12}=1$ and
only one resonance couples to fermions.

\subsection{Electroweak constraints}
\label{modindep}
Precision measurements of electroweak parameters place stringent
constraints on the possibility of VEV for Higgs multiplets other than
the doublet and the \mbox{septet} \mbox{(isospin-3)}. The deviation of the $\rho$
parameter from unity (or, equivalently, the $T$-parameter) constraints
the VEV of the multiplets at tree level:
\beq 
\delta\rho = \frac{M_W^2-M_3^2}{M_3^2} ,
\eeq 
where $M_W$ is the
$W^\pm$ mass and $M_3=M_Z\cos\theta_W$ that of the neutral component
of the $SU(2)$ vector boson multiplet. At tree level, in the case of
$N$ multiplets $H_i$, $i=1,\ldots,N$, with $H_i$ in the
$(2n_i+1)$-dimensional representation of $SU(2)$ with VEV $v_i/\sqrt2$
in the $T_3=m_i$ component (and hypercharge\footnote{ We use the convention $Q=T^3 + Y$ for the electric charge throughout this work.} of $Y(H_i)=-m_i$), we
obtain
\beq
\delta\rho=\frac{\sum_i v_i^2[n_i(n_i+1)-3m_i^2]}{\sum_i 2v_i^2m_i^2},
\label{eq:deltarho}
\eeq
Note that for $gn_i\le4\pi$, a loose requirement for
perturbation theory to hold,  $n_i(n_i+1)-3m_i^2$ vanishes for
$n_i=|m_i|=\frac12$ or $n_i=3$, $|m_i|=2$ only (additional solutions
are found for larger isospin, {\it e.g.}, $n_i=48$, $|m_i|=28$ or
$n_i=\tfrac{361}2$, $|m_i|=\tfrac{209}2$).

Quantum corrections due to additional Higgs bosons can be conveniently studied in terms of the oblique parameters, in particular the well-known $S$ and $T$.  
We do not attempt to study these corrections for arbitrary Higgs representation. The results largely depend on the exact form of the Higgs potential, that we leave unspecified in this work. We refer the readers to the literature for relevant studies.
In the case of multi-Higgs doublets, these corrections are well known~\cite{Grimus:2008nb}; see \cite{Barbieri:2013aza} for a recent discussion.  For the GM model, constraints from EWPD have been analyzed in \cite{Englert:2013zpa}.

In the context of the doublet-septet model, the $S$ and $T$ parameters have only been studied in the special case where the charged Higgs spectrum is taken to be degenerate with the exception of one singly charged Higgs~\cite{Hisano:2013sn}.

\section{Perturbative Unitarity}
\label{sec:unitarity}
Perturbative unitarity of the SM has been famously used to place a
bound of about 700~GeV on the Higgs mass. Lee, Quigg and Thacker (LQT)
\cite{Lee:1977eg} observed that the tree level partial wave amplitudes
for longitudinally polarized $W^+W^-$ scattering grow with the Higgs
mass, so that at large enough mass the amplitudes exceed the unitarity
bound. They also pointed out that in the absence of the Higgs particle
the $J=0,1$ partial wave amplitudes grow with the square of the center
of mass energy $s$, but the exchange of the Higgs particle in the
$s$- and $t$-channels cancels the linear growth with $s$ of these
amplitudes.

\subsection{Sum rule for $W^+_LW^-_L\to W^+_LW^-_L$}
Applying the  LQT argument to the multi-Higgs doublet extension of the
SM gives an alternate derivation of the sum rule \eqref{sumrule1}. It
is the statement that the couplings that appear in the $s$- and $t$-
channel neutral Higgs exchange must add up to those of the SM
contribution in order to cancel the linear growth with $s$ of the
$J=0,1$ partial wave amplitudes.

This suggests a more general approach to the sum rule for any number
of $N$ neutral resonances $h_i$ that couple to $W^+W^-$ with strength
$a_i$, namely
\beq
\label{sumrule3}
\sum_{i=1}^N a_i^2\overset{?}{=} 1.
\eeq 
The constraint \eqref{sumrule3} is stronger than the one in
\eqref{sumrule2}, which must hold in order to obtain the correct $W$
mass in a multi-Higgs model with at least one Higgs multiplet of
isospin~1 or higher. And it is incorrect. In general there are
additional contributions to the $W^+W^-$ scattering amplitude from
$u$-channel exchange of a doubly charged component of the multiplet to
which $h_i$ belongs. 

\begin{figure}
  \centering
 \subfloat{\includegraphics[width=0.9\textwidth]{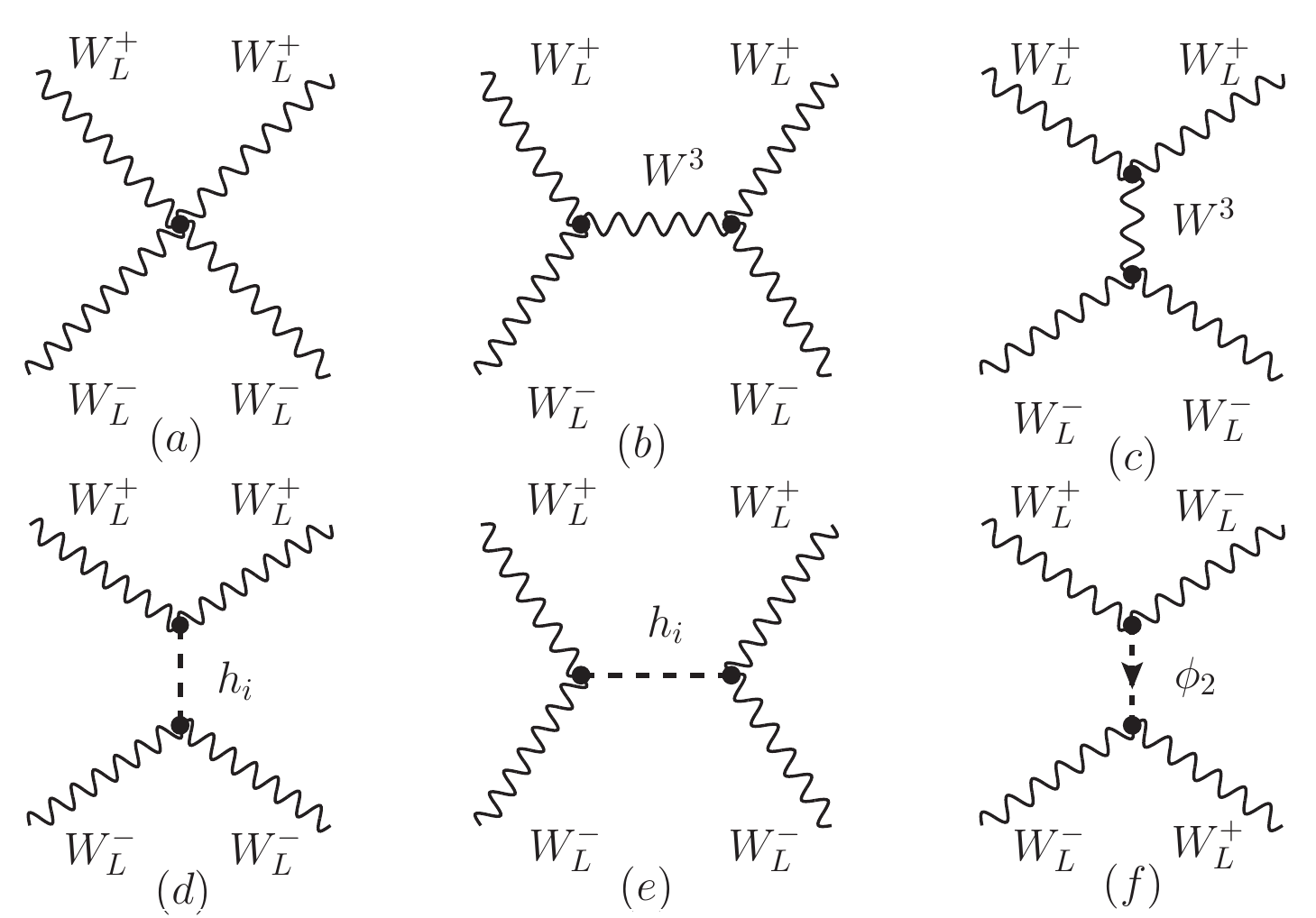}} 
\caption{$W^+W^-$ scattering in the presence of a generic Higgs sector. Diagrams a-c refer to the scattering in the Higgsless standard model, with $W^3$ collectively denoting the exchange of the $Z$ boson and the photon. The growth with the center of mass energy of these contributions can be cancelled by neutral (d, e) or doubly charged (f) Higgses.}
 \label{fig:WWscattering}
\end{figure}

To see that this is the case we compute the  amplitude for
longitudinal $W^+W^-$ scattering including contributions from a
neutral Higgs with arbitrary coupling $gM_Wa$ to $W^+W^-$ and of a doubly charged
complex scalar with arbitrary coupling $gM_Wb$ to $W^+W^+$ (plus hermitian
conjugate). Only the $J=0$ and 1 partial wave amplitudes
exhibit linear growth with $s$, and the coefficient of the term exhibiting linear
growth  is common to both amplitudes and proportional to 
\beq
\label{eq:sgrowthcoef}
\kappa_1=4-3(M_3/M_W)^2-a^2+4b^2 .
\eeq 
Here, the first two terms come from the pure gauge sector,
Fig.~\ref{fig:WWscattering}a-c.  The second term arises when the
exchanged neutral gauge boson is massive,
Fig.~\ref{fig:WWscattering}b and ~\ref{fig:WWscattering}c.  The third term is the
contribution of the neutral Higgs exchange in both $s$- and
$t$-channels, Fig.~\ref{fig:WWscattering}d and ~\ref{fig:WWscattering}e.  The last term
comes from the $u$-channel exchange of a doubly charged scalar,
Fig.~\ref{fig:WWscattering}f. In Eq.~\eqref{eq:sgrowthcoef} the
terms $a^2$ and $b^2$ should be replaced by a sum over squares of
couplings, $\sum a_i^2$ and $\sum_r b_r^2$, when more than one neutral
or doubly charged states are present.
The correct
version of the sum rule in Eq.~\eqref{sumrule3} reads
\beq
\label{sumrule4}
\sum_i a_i^2-4\sum_r  b_r^2 = 4-3(M_3/M_W)^2~. 
\eeq 

Following LQT we can also obtain an upper limit on a combination of
the masses $M^0_i$ and $M^{++}_r$ of the neutral and doubly charged
Higgses. We quickly review the LQT computation. They consider (in the
SM) first the limit $M_h\gg M_W\sim M_Z$ of the $J=0$ partial wave
scattering amplitude for $W_L^+W_L^-\to W_L^+W_L^-$, followed by the  large CM
energy limit. They find the following expression for (the finite piece of) the $s$-wave scattering amplitude
\beq
a_0(W_L^+W_L^-\to W_L^+W_L^-)\xrightarrow[s\gg M_h^2\gg M_W^2]{} -\frac{G_FM_h^2}{4\pi\sqrt2},
\eeq
from which the condition $\big| a_0(W_L^+W_L^-\to W_L^+W_L^-)\big|\le 1$ gives
$M_h^2\le 4\pi\sqrt2/G_F$. LQT derive a slightly better bound by
performing a coupled channel analysis including also $ZZ$, $Zh$ and $hh$
scattering. 

Below we use a slightly more constraining condition that follows from
unitarity, $\big|\text{Re}\big[ a_0(W_L^+W_L^-\to W_L^+W_L^-)\big]\big|\le
1/2$. The same procedure gives, for the more general case considered
here,
\beq
\label{eq:massboundWW}
\sum_i ( a_iM^0_i)^2+2\sum_r (b_rM^{++}_r)^2 \le  \frac{2\pi\sqrt2}{G_F}\approx0.5~\text{TeV}^2~.
\eeq
To obtain this bound the limit of small $M_{W,Z}$ is taken first at constant
Higgs masses, and only then the large $s$ limit is taken. The
contribution of the $ a_1\approx1$, $M^0_1\approx 126$~GeV Higgs to the
bound is negligible (which is consistent with the approximation of
neglecting the similarly small masses $M_W$ and $M_Z$).

\subsubsection{Examples}
For explicit examples, we first consider the simpler case of an
$SU(2)$ gauge theory with a single Higgs field $\Phi$ with
isospin~$(2n+1)$. We assume the VEV of the field, $v/\sqrt2$, is in
the $T^3=m$ component of the multiplet with hypercharge $Y=-m$. Let
\beq
\Phi-\VEV{\Phi}=\sum_k \phi_k |k\rangle,
\eeq
where $|k\rangle$ is a normalized $2n+1$ dimensional vector in isospin
space, with $T^3|k\rangle=k|k\rangle$. Then the lagrangian contains a
term 
\beq
\mathcal{L}\supset \frac{g^2v}{\sqrt2}\eta^{\mu\nu}
\left[A\left(W^+_\mu W^+_\nu\phi_{m-2}+
W^-_\mu W^-_\nu\phi_{m-2}^*\right)
+B\left(W^+_\mu W^+_\nu\phi_{m+2}^*+
W^-_\mu W^-_\nu\phi_{m+2}\right)\right],
\eeq
where\footnote{ Our normalization conventions are $[T^a,T^b] = i\epsilon^{abc}$ with $\epsilon^{123}=1$, and $[T^+,T^-] = T^3$.
}
\beq
\begin{aligned} \label{eq:AB}
A=A(n,m) &=\VEV{m|T^+T^+|m-2}=\VEV{m-2|T^-T^-|m}\\
&=\frac{1}{2}\sqrt{\left[n(n+1)-(m-2)(m-1)\right]\left[n(n+1)-m(m-1)\right]}\,,\\
B=B(n,m)&=A(n,m+2) =\VEV{m+2|T^+T^+|m}=\VEV{m|T^-T^-|m+2}\\
&=\frac{1}{2}\sqrt{\left[n(n+1)-m(m+1)\right]\left[n(n+1)-(m+1)(m+2)\right]}\,.
\end{aligned}
\eeq Note that $A=0$ for $m=-n$ and $m=-n+1$, and $B=0$ for $m=n$ and
$m=n-1$, so these coefficients automatically account for the absence
of charged states with disallowed isospin components, {\it
  i.e.}, $m=n+2$, $n+1$, $-n-1$ or $-n-2$. Of course, $A=B=0$ in the
case $|m|=n=\tfrac12$. Using $M_W^2=\tfrac14g^2v^2C$ where
$C=2[n(n+1)-m^2]$ as per Eq.~\eqref{eq:cgiven}, $M_3^2=g^2v^2m^2$,
$a=\sqrt{C}$ from Eq.~\eqref{eq:genrlzdWWh} and $\sum_i b_i^2=2(A^2+B^2)/C$,
one finds $\kappa_1=0$ in Eq.~\eqref{eq:sgrowthcoef}.  Note that for
all cases other than $n=\tfrac12$ (the SM) and $n=1$ with $Y=0$ (the
prototypical triplet Higgs model), the contribution of the doubly
charged Higgs particle is crucial to insure perturbative unitarity at
high energies. The generality of this result is remarkable. Note for
example, that for integer $n$ with $Y=0$ the pattern of symmetry
breaking is $SU(2)\to U(1)$ so that $W^3$ remains exactly massless. In
this case the vanishing of $\kappa_1$ results from the cancellation of
the first, third and fourth terms in Eq.~\eqref{eq:sgrowthcoef}. The second term is absent because  $W^3$ is massless.

The bound on the masses in
\eqref{eq:massboundWW}, assuming for simplicity that the two doubly charged states are mass-degenerate with masses $M^{++}$, gives $$C(M^0)^2+4\frac{A^2+B^2}{C}(M^{++})^2\le
2\pi\sqrt2/G_F\approx0.5~\text{TeV}^2.$$ 
For a concrete and pertinent example,
take $n=3, m=2$; then $16(M^0)^2+\tfrac{15}2(M^{++})^2\le0.5~\text{TeV}^2$. This
is very constraining: it gives $M^0\le177$~GeV, $M^{++}\le259$~GeV,
and if the masses are comparable, $M^0\approx M^{++}\le146$~GeV.

One can readily generalize above discussion to the realistic case with
gauge group $SU(2)_L \times U(1)_Y$. Electric charge conservation requires
the VEV in the electrically neutral component,
$Q=T^3+Y=0$, fixing the hypercharge of the Higgs multiplet,
$Y_\phi=-m$. This implies that the  $Z$-boson, $Z=\cos \theta_W W^3-\sin
\theta_W B$, has   mass given by 
\beq M_Z=\sqrt{g^2+g'^2} ~m v =\frac{g }{\cos\theta_W} ~m v~.
\label{eq:zmass}
\eeq 
Here, $g$ and $g'$ are the gauge couplings for $SU(2)_L$ and $U(1)_Y$
respectively, while the electroweak angle is defined in the standard
way, $\cos \theta_W=g/\sqrt{g^2+g'^2}$.  The sum rule obtained by
setting $\kappa_1$ in \eqref{eq:sgrowthcoef} to zero is automatic,
provided one  substitutes 
\beq 
M_3^2\to g^2 m^2 v^2=M^2_Z\cos^2\theta_W .
\eeq

The latter substitution can be understood by noting that the only
differences in the computation for longitudinal $W$ scattering through
$Z$ exchange in a theory with $U(1)_Y$ come from different couplings
and intermediate vector masses. One can show by a straightforward
computation, that while the part of $W_L$ scattering amplitude that
grows as $s^2$ cancels identically in the gauge sector, the linear
piece is given as follows
\begin{align}
\mathcal{M}_{lin}&=-g^2 s \bigg[\frac{3 \cos \theta-1}{2 M_W^2} 
+\cos^2\theta_W \( \frac{\cos\theta M_Z^2}{4M_W^2}
+\frac{ M_Z^2~(3+\cos\theta)-16M_W^2\cos\theta}{8 M^4_W}\)
\nn \\&-\sin^2\theta_W\frac{2 \cos\theta}{M^2_W}\bigg ]
=g^2 s ~(1+\cos\theta)~\frac{•4 M^2_W-3 M^2_Z\cos^2\theta_W}{8 M^4_W}~.
\end{align}
The first term in the first line corresponds to the contact
interaction, the second and third terms, with coefficient
$\cos^2\theta_W$ are the $s$-and $t$-channel $Z$ exchange,
respectively, and the fourth term, proportional to $\sin^2\theta_W$,
is from photon exchange.

Adding the contributions of the neutral and doubly charged Higgs bosons
leads to the following sum rule in a $SU(2)_L \times U(1)_Y$ -
invariant theory with a generic Higgs sector 
\beq 
4 -3 \frac{M_Z^2 \cos^2\theta_W}{M_W^2} - a^2+4 b^2= \(4-\frac{3}{1+\delta
  \rho}\) - a^2+4 b^2=0~, 
\eeq 
where $\delta\rho\equiv\frac{M^2_W}{M^2_Z\cos^2\theta_W}-1$, is
given by the right hand side of \eqref{eq:deltarho}.

\subsection{Sum rule from $Z_LZ_L\to W_L^+W_L^-$}

We have shown above that a charged scalar $u$-channel exchange in a theory with a generic Higgs sector affects in a non trivial way the sum rules that the neutral Higgs boson couplings to the vector mesons should satisfy. 
Here we also derive yet another non-trivial sum rule by demanding perturbative unitarity in the $Z_LZ_L\to W_L^+W_L^-$ channel.
Again we start by considering generic Higgs couplings.  We denote the
coupling of the neutral Higgs to $W^+W^-$ and $ZZ$ by $gM_Wa$ and $gM_Wd/2$ respectively, 
and the coupling of a singly charged Higgs to $ZW^+$ and its hermitian
conjugate by $gM_Wf$ (we assume $f$ is real).  Note that, in the
spirit of a model independent analysis,  we have kept
the couplings of the neutral Higgs to $WW$ and $ZZ$ independent,
although one may expect $d=a /\cos^2\theta_W$ by custodial symmetry.  
Unlike in the $ W_L^+W_L^-\to W_L^+W_L^-$
scattering case, only the $J=0$ partial wave amplitudes exhibit the
linear growth in $s$ proportional
to
\begin{equation}
\label{sumruleZZ}
	\kappa_{2} = \cos^2\theta_W M_Z^4/M_W^4 + f^2-ad~.
\end{equation}
The first term arises from the four-point gauge interactions, as well as from the $t$- and $u$- channel
$W^\pm$ exchange, while the second and third terms are contributed by
the singly charged and neutral Higgs bosons in the ($t$-, $u$-) and $s$-
channels, respectively. The above sum rule must be treated with
care. The reason for this is that  one
combination of  the singly charged Higgs bosons is  eaten by the
$W^\pm$. It is then necessary to eliminate the fake contribution of the
goldstone combination to the $ZZ\to WW$ scattering; see below for an
explicit example.  In a generic case
of arbitrary number of neutral and singly charged scalars, we obtain
the following sum rule
\begin{equation}
\label{eq:sumruleADF}
	\cos^2\theta_W M^4_Z / M_W^4+\sum_r f_r^2 - \sum_i
        a_i d_i = 0,
\end{equation}
where only the \textit{physical} states are understood to contribute
to the sum rule. If one insists on $d_i=a_i/\cos^2\theta_W$, as one
may expect by custodial symmetry, then this is
$\sum_ia_i^2-\cos^2\theta_W\sum_r f_r^2 = (\cos\theta_W M_Z/M_W)^4$. One  may combine this result with
the sum rule in Eq.~\eqref{sumrule4}, and use 
$\delta\rho\ll1$ to obtain a connection between singly and doubly
charged Higgs resonances, $\cos^2\theta_W \sum_r f_r^2=4\sum_r
b_r^2$. An immediate consequence is that in multi-higgs doublet models
the couplings of charged higgs particles to $WZ$ vanish.

The subleading, $s$-independent piece leads to the constraints on the charged and neutral Higgs masses from perturbative unitarity; again, we can obtain LQT-like mass bounds from the requirement
  that the $J=0$ partial wave respect unitarity, $|\text{Re}(a_0)|<1/2$:
\begin{equation}
\label{massboundZZ}
\sum _i a_i d_i (M^0_i)^2+2 \sum_r f_r^2(M^+_{r})^2 <\frac{4\pi\sqrt2}{\cos^2\theta_WG_F}\approx 1.3~\text{TeV}^2.
\end{equation}

\subsubsection{Example: A Single Electroweak Multiplet}

We illustrate these results with a model consisting of a single Higgs
field belonging to an arbitrary representation of the
$SU(2)_L \times U(1)_Y$ gauge group. Consider a multiplet $\phi$
having isospin $2 n+1$ and vacuum expectation value $\langle
\phi_m\rangle =v/\sqrt{2}$ on the component with $T^3=m$ (hence
$Y=-m$). 
For general $n$, there are two
singly charged Higgses, $\phi^*_{m+1}$ and $\phi_{m-1}$. Expanding out
the kinetic term, one finds their couplings to the gauge bosons \beq
\begin{split}
  \mathcal{L} &\supset 
    \frac12\eta^{\mu\nu}\left[g^2W^+_\mu W^-_\nu\left(n(n+1)-m^2\right)
         +(g^2+g^{\prime2})Z_\mu Z_\nu m^2\right](v+\phi_m)^2\\
  &\qquad + \frac{v\cos\theta_W}{\sqrt{2}}~ \bigg [ Z^\mu W^{+}_\mu
  \(D\phi_{m-1}+E\phi^*_{m+1}\) + h.c.\bigg ],
\end{split}
\eeq
where $g^\prime$ is the hypercharge gauge coupling and
\begin{align}
D&=D(n,m)= F \big[2 (g^2+g'^2)m-g^2 \big ]~,\nn\\
E&= E(n,m)=G \big[2 (g^2+g'^2)m+g^2 \big ]\nn~.
\end{align}
Here we have defined
\begin{align}
F&=F(n,m)=\sqrt{\tfrac12\big(n(n+1)-m(m-1)\big)}~,\nn\\
G&=G(n,m)=F(n,m+1)=\sqrt{\tfrac12\big(n(n+1)-m(m+1)\big)}\nn~.
\end{align}

One linear combination, $\chi^-$, of $\phi_{m-1}$ and
$\phi_{m+1}^\ast$ is eaten up by $W^+$ while the orthogonal combination is
the physical charged Higgs, $\phi_P^-$:
\beq \chi^-=\frac{F\phi_{m-1}-G\phi^*_{m+1}}{\sqrt{F^2+G^2}}, \quad \phi_P^- =
\frac{G\phi_{m-1} + F\phi^*_{m+1}}{\sqrt{F^2+G^2}}.  
\eeq 
The physical singly charged Higgs couples to the $WZ$ pair through the following Lagrangian operator
\begin{equation}
\mathcal{L}_{phys} \supset
\frac{v\cos\theta_W}{\sqrt{2}}\,
\frac{4m\,F \,G(g^2+g^{\prime2})}{\sqrt{F^2+G^2}}\, 
W^+_\mu Z^\mu \phi_P^-+ \hc
\end{equation}
The scalar couplings that enter into the sum rule~\eqref{sumruleZZ}
are thus identified,
\begin{equation}
\begin{split}
	gM_W a &= g^2v\left[n(n+1)-m^2\right],\\
	gM_W d &=2 (g^2+g^{\prime2})v\,m^2,\\
	(gM_W f)^2 &=\(\frac{v\cos\theta_W}{\sqrt{2}}\,
\frac{4m\,F \,G(g^2+g^{\prime2})}{\sqrt{F^2+G^2}}\)^2.
\end{split}
\end{equation}
As a consistency check, note that $\tilde f^2$ vanishes for the standard doublet Higgs representation
$n=|m|=1/2$. 

Of course, as can be straightforwardly checked, these coefficients satisfy the sum rule in
Eq.~\eqref{sumruleZZ} automatically, for any choice of $n$ and
$m$. The perturbative unitarity bound \eqref{massboundZZ} on the masses of extra charged Higgs bosons is
remarkably strong, as long as the multiplets to which they belong significantly contribute to electroweak symmetry breaking.

The limits on the singly and doubly charged Higgs masses that follow from unitarity are explored for a few interesting cases in the end of this section.

\subsubsection{Generalization}
\label{sec:couplings}
For completeness, we extend the analysis of the previous example to the case with many electroweak multiplets, each of which can belong to an arbitrary representation. 
The normalized goldstone mode now reads
\begin{equation}
\begin{split}
	\chi^- = \frac{\displaystyle\sum_iv_i\left(F_i\phi_{m-1}^{(i)}-G_i\phi_{m+1}^{(i)\ast}\right)}{\sqrt{\displaystyle\sum_j v_j^2\left(F_j^2+G_j^2\right)}}
	= \frac{g}{\sqrt{2}M_W}\sum_i v_i\sqrt{F_i^2+G_i^2}\, \chi^{(i)-},
\end{split}
\end{equation} 
where $\chi^{(i)}$ is the would-be goldstone combination contributed by the
$i^{\text{th}}$ multiplet and we have defined, $F_i\equiv F(n_i,m_i)$ and $G_i\equiv G(n_i,m_i)$.

Having identified the goldstone mode, one can straightforwardly work out all the physical singly charged modes and their couplings to gauge bosons.  
However, the expressions for physical modes in a generic case is not particularly illuminating.  

Instead we consider a specific example with an extended Higgs sector, consisting of of the usual electroweak doublet with hypercharge 1/2 and an electroweak septet with hypercharge 2, \textit{i.e.} $(n_1,\,m_1) = (1/2,\,-1/2)$ and $(n_2,\,m_2)=(3,\,-2)$; we will briefly touch on the phenomenology of this theory below. 
The goldstone mode in this case reads
\begin{equation}
	\chi^- = \frac{g}{\sqrt{2}M_W}\left(- v_1 G_1 \phi^{(1)\ast}_{1/2} + v_2 F_2 \phi^{(2)}_{-3} - v_2 G_2 \phi^{(2)\ast}_{-1}\right),
\end{equation}
while the two orthogonal singly charged Higgses are
\begin{equation}
\begin{aligned}
	\phi_1^- &= \frac{v_2 F_2}{\sqrt{v_1^2 G_1^2 + v_2^2F_2^2}} \phi^{(1)\ast}_{1/2} + \frac{v_1 G_1}{\sqrt{v_1^2 G_1^2 + v_2^2F_2^2}} \phi^{(2)}_{-3},\\
	\phi_2^- &= \frac{g}{\sqrt{2}M_W}\left(\frac{v_1v_2G_1 G_2}{\sqrt{v_1^2 G_1^2 + v_2^2F_2^2}} \phi^{(1)\ast}_{1/2} 
		- \frac{v_2^2 F_2 G_2}{\sqrt{v_1^2 G_1^2 + v_2^2F_2^2}} \phi^{(2)}_{-3}
		-\sqrt{v_1^2 G_1^2 + v_2^2F_2^2}\,\phi^{(2)\ast}_{-1}\right).
\end{aligned}
\end{equation}

In general, the two neutral CP-even Higgses, and the two singly charged Higgses mix among themselves.  
The weak eigenstates are related to the physical fields via mixing angles, which we define as follows:
\begin{equation}
	\begin{pmatrix}\phi^{(1)}_{-1/2}\\ \phi^{(2)}_{-2} \end{pmatrix}
	= \begin{pmatrix}c_\alpha & s_\alpha\\ -s_\alpha & c_\alpha\end{pmatrix}
	\begin{pmatrix}h\\ H \end{pmatrix},
	\quad
	\begin{pmatrix}\phi^-_1\\ \phi^-_2 \end{pmatrix}
	= \begin{pmatrix}c_\gamma & s_\gamma\\ -s_\gamma & c_\gamma\end{pmatrix}
	\begin{pmatrix}h^- \\ H^- \end{pmatrix}.
\end{equation}
Defining $t_\beta = v_1/4v_2$, the couplings of the \textit{physical} states relevant for the mass bound in equation~\eqref{massboundZZ} are given by the following expressions
\begin{equation}
\label{eq:physcouplings}
\begin{aligned}
	a_h &= \frac{M_W^2}{M_Z^2}d_h = c_\alpha s_\beta - 4s_\alpha c_\beta,\\
	a_{h^\prime} &= \frac{M_W^2}{M_Z^2}d_{h^\prime} = s_\alpha s_\beta + 4c_\alpha c_\beta,\\
	b &= \frac{\sqrt{15}}{2} \cos\beta ~,\\
	f_h &= -\frac{M_Z}{M_W}\frac{c_\beta(5\sqrt{3}s_\beta c_\gamma + 3\sqrt{5}s_\gamma)}{\sqrt{3+5s_\beta^2}},\\
	f_{h^\prime} &= \frac{M_Z}{M_W}\frac{c_\beta(3\sqrt{5} c_\gamma - 5\sqrt{3}s_\beta s_\gamma)}{\sqrt{3+5s_\beta^2}}.
\end{aligned}
\end{equation} 
The sum rule \eqref{eq:sumruleADF} is satisfied with
$\cos^2\theta_W\sum_r f_r^2= 15c_\beta^2$. The sum rule  $\cos^2\theta_W \sum_r f_r^2=4\sum_r
b_r^2$ then gives the correct coupling of the only doubly charged higgs to $WW$.

\subsubsection{Bounds on Extra Higgs Masses from Unitarity}
\label{subsec:massbound}
We close the present subsection by an illustration of the unitarity constraints on masses of various extra Higgs bosons present in the doublet-septet model. The parameters $a,b,d,f$, entering the perturbative unitarity conditions depend on the angles $\alpha$, $\beta$ and $\gamma$; hence the bounds on physical masses, such as $m_{h'}$ and $M_{h^{++}}$, implied by the sum rules
largely depend on the mixing angles (as well as other SM parameters and the mass of the lighter neutral Higgs, that is assumed to be $m_h=126$~GeV). One
way to visualize the constraints imposed by unitarity is to fix
various values for the two angles and explore the allowed regions for
the extra scalar and charged Higgses present in the theory. We start by exploring the bounds from $WW$ scattering, eq. \eqref{eq:massboundWW}.

At $\alpha=0$, $\beta=\pi/2$, there is no VEV for the septet nor mixing
between the neutral CP-even states. This is like the single Higgs
doublet case and hence perturbative unitarity does not lead to any
constraints on the extra Higgs masses (there are really no extra
Higgses, states that couple directly to pairs of vectors bosons).
\begin{figure}
  \centering
\subfloat{\includegraphics[width=0.42\textwidth]{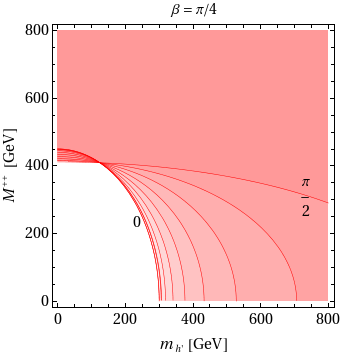}}
\subfloat{\includegraphics[width=0.42\textwidth]{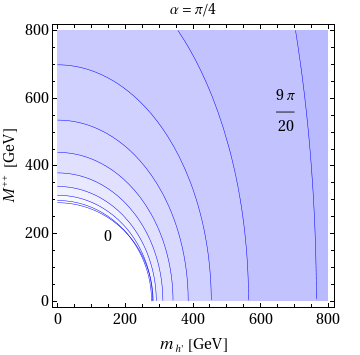}}
\caption{Allowed regions for the heavy neutral and doubly charged
  Higgs boson masses for various values of mixing angles in the
  doublet-septet model.  The excluded regions are shaded. On the left figure $\beta$ is fixed at $\pi/4$ and $\alpha$ is increased from the left to right curves in steps of $\pi/20$. On the right figure, $\alpha$ is fixed at $\pi/4$ and $\beta$ increases from left to right in steps of $\pi/20$.}
  \label{fig:wwmm}
\end{figure}
The case $\alpha=\beta=\pi/2$ again corresponds to having EW symmetry
broken solely by the doublet Higgs. The doubly charged state from
the septet therefore does not couple to the vector bosons, so that its
mass is not constrained by unitarity. On the other hand, the mixing
between the neutral states from both multiplets is maximal, so that
$h$ is actually purely the neutral component of the septet. This
does lead to a constraint $m_{h'}\lsim 900$~GeV but since the
$126$~GeV Higgs does not couple to $W$ in this case, the situation is clearly
unphysical. The opposite (and again unphysical) case of $\alpha=\pi/2,~
\beta=0$, where the EW symmetry is broken purely by the septet VEV,
leads to $M^{++}\lesssim 300$~GeV while leaving $m_{h'}$ unconstrained. Two
intermediate cases are shown in Fig.~\ref{fig:wwmm}. 
The perturbative unitarity bounds on the extra Higgs
masses for this case, as one can see, are quite stringent. 

\begin{figure}
  \centering
  \subfloat[]{\label{fig:uni1}\includegraphics[width=0.42\textwidth]{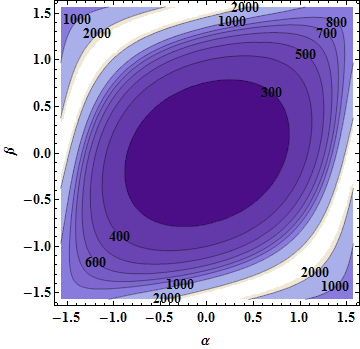}}\hspace{2mm}
  \subfloat[]{\label{fig:uni2}\includegraphics[width=0.42\textwidth]{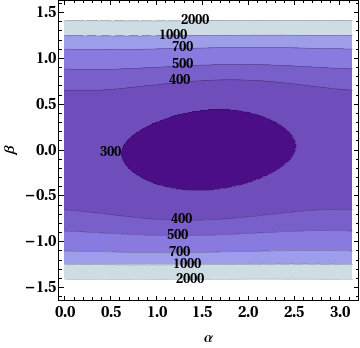}}
  \caption{Conservative perturbative unitarity upper bounds on $m_{h'}$ (\ref{fig:uni1}, for $M^{++}=0$) and $M^{++}$ (\ref{fig:uni2}, for $m_{h^\prime}=0$) as a function of mixing angles in the doublet-septet
model. The bounds are periodic in $\alpha$ and we have centered the figures around the value of $\alpha$ where the bound is strongest.}
  \label{fig:uni}
\end{figure}

Conservative bounds on the heavy Higgs masses, implied by perturbative unitarity in the doublet-septet model are shown in Fig.~\ref{fig:uni}, assuming the lighter neutral Higgs mass $m_h=126$~GeV.  In Fig.~\ref{fig:uni1} (\ref{fig:uni2}) the positive definite contribution of $M^{++}$ ($m_{h^\prime}$) in equation~\eqref{eq:massboundWW} has been ignored.  In both figures the white regions correspond to small couplings $a_{h^\prime}$ and $b$, implying therefore very weak bounds on the Higgs masses.

\begin{figure}
  \centering
\subfloat{\label{fig:wzmm3}\includegraphics[width=0.42\textwidth]{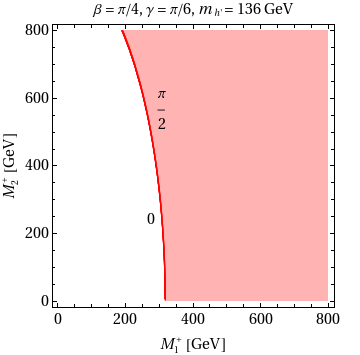}}
\subfloat{\label{fig:wzmm4}\includegraphics[width=0.42\textwidth]{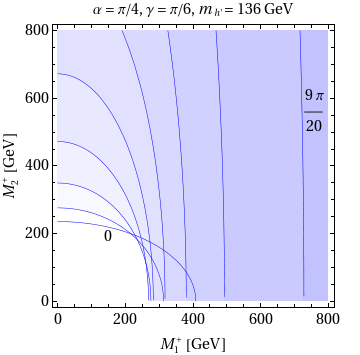}}\\
\subfloat{\label{fig:wzmm3b}\includegraphics[width=0.42\textwidth]{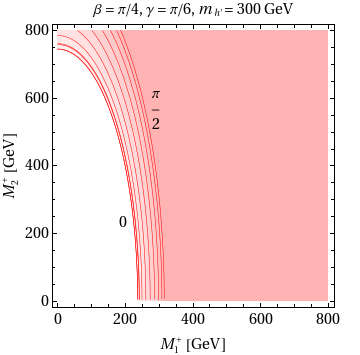}}
\subfloat{\label{fig:wzmm4b}\includegraphics[width=0.42\textwidth]{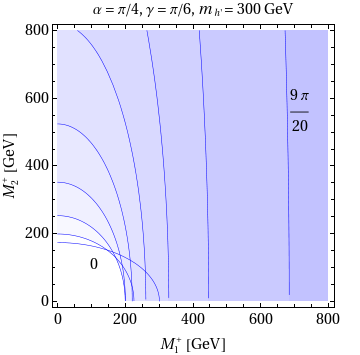}}
  \caption{Allowed regions for the singly charged Higgs boson masses
    for various values of mixing angles in the doublet-septet model. The heavy neutral Higgs masses have been taken to be 136 GeV (top row) and 300 GeV (bottom row). Shaded regions are excluded. On the left column $\beta$ is fixed at $\pi/4$ while $\alpha$ is increased from the left to right curves in steps of $\pi/20$. On the right column $\alpha$ is fixed at $\pi/4$ and $\beta$ increases from left to right in steps of $\pi/20$. For all of the above figures $\gamma$ has been fixed at $\pi/6$.}
  \label{fig:wzmm}
\end{figure}

Similar bounds on the masses of the two singly charged Higgses, $M_1^+$ and 
$M_2^+$, implied by perturbative unitarity in the $WZ$ channel are
shown in Fig.~\ref{fig:wzmm}. The value for the singly charged Higgs mixing angle $\gamma$ is taken to be $\pi/6$ for all cases. The bound is obtained for two different values of the heavy neutral Higgs mass, $m_{h^\prime} = 136$, and 300 GeV respectively. 
In each case, one of the  mixing angles $\alpha$ and $\beta$ is kept constant while the other is varied. Again, for the case that the septet takes appreciable part in the electroweak symmetry breaking ($\beta$ is not close to $\pi/2$), at least one of the charged Higgs bosons is bound to be relatively light. 

\subsection{The sum rule for $W^+_LW^-_L\to t \bar t$}
\label{sec:fermionsumrule}
\begin{figure}
  \centering
 \subfloat{\label{fig:rateKM}\includegraphics[width=1.0\textwidth]{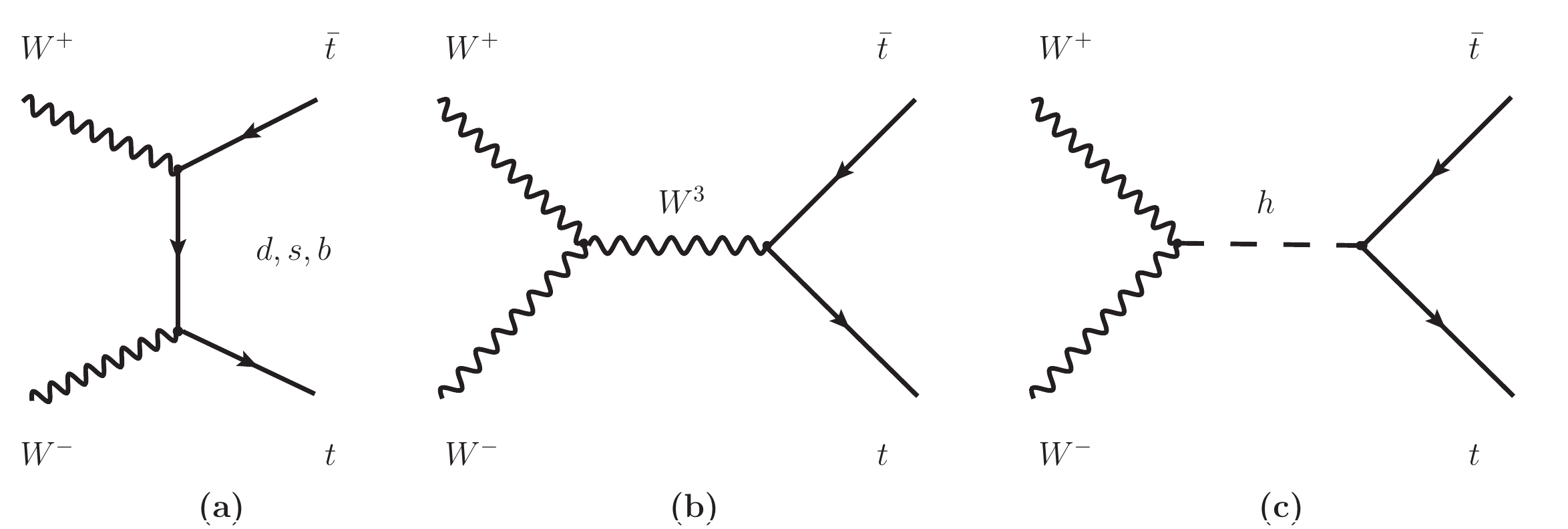}} 
\caption{$W^+W^-\to t\bar{t}$ scattering in the presence of a generic Higgs sector. $W^3$ in diagram b denotes the exchange of the $Z$ boson and the photon. Similarly, the $h$ in diagram c stands for a generic neutral Higgs boson.}
 \label{fig:WWtt}
\end{figure}

A sum rule for $W^+_LW^-_L\to t \bar t$ is useful in constraining the
couplings of the neutral scalars to the top-quark, which contribute to
the amplitudes for production of these states in $gg$-fusion and for
the decay into $\gamma\gamma$ and $\gamma Z$. The Feynman diagrams
contributing to $W^+_LW^-_L\to t \bar t$ are shown in Fig.~\ref{fig:WWtt}. The
growth with one power of $s$ cancels among diagrams \ref{fig:WWtt}(a)
and~\ref{fig:WWtt}(b), resulting in a leading contribution that grows as $\sqrt{s}$.  

We denote the coupling of the $i^{\text{th}}$ neutral, physical CP-even
scalar to $\bar t t$ by $\lambda_i/\sqrt2$, so that
$\lambda=gm_t/\sqrt2M_W$ in the SM as usual. Insisting that the growth
with $\sqrt{s}$ is cancelled by the Higgs exchange diagram \ref{fig:WWtt}(c), we derive the sum rule
\begin{equation}
\label{sumruletop}
\frac{gm_t}{\sqrt{2}M_W}-\sum_i  a_i\lambda_i=0 ,
\end{equation}
where, as before, we denote by $a_i = (R^T)_{ij} \tilde a_j$ and $ \lambda_i=(R^T)_{ij}\tilde  \lambda_j$ the
couplings of the physical, mass eigenstate Higgs to $W^+W^-$ and $\bar
tt$, respectively, where $R$ is an orthogonal matrix, transforming weak eigenstates into the physical ones.
The above equation has a simple interpretation. The
sum of the couplings of the various scalars to the top quark has to be
such that when the $i^{\text{th}}$ scalar is replaced by its expectation value,
$v_i$, it gives the top quark mass term: $m_t=\sum
\tilde{\lambda}_iv_i/\sqrt2$. One can see this by writing the above sum rule in the weak eigenbasis,
\begin{equation}
\label{pre-sumruletop}
\frac{gm_t}{\sqrt{2}M_W}-\sum_i \tilde a_i\tilde \lambda_i=0.
\end{equation}
Using $v_i=\tilde a_i(2M_W/g)$, one can see that this exactly
reproduces the expression for the top quark mass.  The sum rule is
more conveniently written as per  Eq.~\eqref{eq:lag-prime} in terms of the deviation of each Yukawa
coupling from the SM value, $\lambda_i = (gm_t/\sqrt2M_W)c_{ti}$, thus
 \begin{equation}
\label{full-sumruletop}
\sum_i a_i c_{ti}=1.
\end{equation}

Again, only the neutral Higgs from an $SU(2)$-doublet may couple
to $\bar tt$. For example, if only the first Higgs is from a doublet,
then $ \lambda_i=(R^T)_{i1}\tilde\lambda_1$, and $\sum
\lambda_i^2=\tilde\lambda_1^2$, or $\sum c^2_{ti}=1/a_1^2\ge1$.

The sum rule \eqref{sumruletop} has immediate phenomenological
implications. It is saturated by any Higgs that has SM-like
couplings to both $W^+W^-$ and $\bar t t$. It follows that either one or the
other of these couplings for additional Higgses must vanish or there
must be at least two additional Higgses with canceling
contributions. That is, if a second Higgs-like resonance is discovered
with near SM-like couplings, then a third one must also
exist. Moreover, both of these resonances would have SM-like cross
section for $gg\to h\to WW$, and one of them would have enhanced decay
rate into $\gamma\gamma$ (since the $t$-loop and the $W$-loop contributions
would interfere constructively). 

Similar sum rules apply to the rest of quarks and all charged
leptons. If the 126 GeV Higgs is observed not to decay (or have
suppressed decays) to any one quark or charged lepton it follows
immediately from the sum rule that there must be at least another
CP-even neutral Higgs.

\section{Model Independent Analysis}
\label{model-indep}

\subsection{Neutral Higgses}

\begin{figure}
  \centering
 \includegraphics[width=0.5\textwidth]{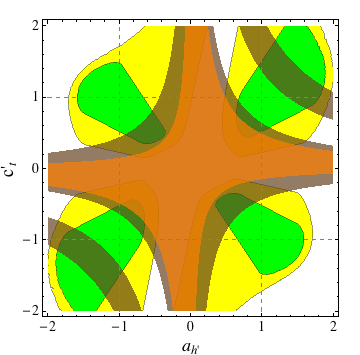}
\caption{Region in the $a_{h^\prime}$-$c_t^\prime$ parameter space
  consistent with the unitarity constraint of Eq.~\eqref{sumruletop}
  assuming $a_h$ and $c_t$ are determined from the 126 GeV Higgs data.
  The orange and brown regions are compatible with 126 GeV Higgs data
  at the 68\% and 95\% CL, respectively. For reference, the figure has been
  superimposed on the fit in Fig.~\ref{fig:136} to the CMS data
  suggesting a Higgs resonance at 136~GeV.}
 \label{fig:modelindependentfit}
\end{figure}
 
In general, the couplings of any extra neutral Higgs, $h^\prime$, are related to the couplings
of the $126$ GeV state $h$ via unitarity constraints of Section~\ref{sec:unitarity}.  In
the following, we assume that there is only one additional
neutral CP-even Higgs, and possibly some charged Higgs bosons.
Here we focus on the sum rule involving the couplings to top
quarks and vector-bosons, Eq.~\eqref{full-sumruletop},   since
it doesn't include the charged Higgs couplings and is therefore the most robust:
\begin{equation}
\label{eq:ttwwsumrule}
	a_{h^\prime} c_t^\prime = 1 - a_h\,c_t.
\end{equation}
\begin{figure}
  \centering
\includegraphics[width=0.45\textwidth]{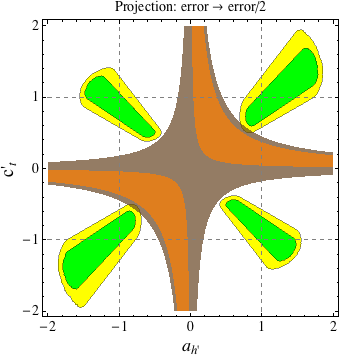}
 \includegraphics[width=0.45\textwidth]{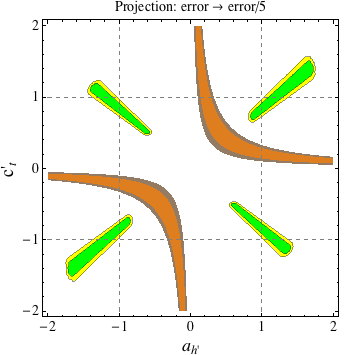}
\caption{Projection of the viable parameter space assuming all the errors are reduced by a factor of 2 (left plot) and 5 (right plot).}
 \label{fig:projections}
\end{figure}
By fitting $a_h$ and $c_t$ to the existing 126 GeV Higgs data, we 
determine the allowed values for the quantity $a_h c_t$ at  68\% and
95\% CL.  Using the sum rule \eqref{eq:ttwwsumrule}, the allowed
region can be mapped onto the $a_{h^\prime}$ -- $c_t^\prime$
plane. The result is shown in Fig.~\ref{fig:modelindependentfit},
where regions in the $a_{h'}$ -- $c_t^\prime$ plane  compatible with the 126 GeV Higgs data at the    68\%  and
95\% CL   are depicted in orange and brown, respectively. This is
superimposed on Fig.~\ref{fig:136} of the parameter estimation in
light of the CMS 136~GeV higgs-like resonance data.  One can see that
only a small portion of $a_{h^\prime}$ -- $c_t^\prime$ parameter
space, allowed by current CMS measurements on the $136$ GeV resonance
is actually consistent with the 126 GeV Higgs data.  It is important
to note that the bound from Eq.~\eqref{eq:ttwwsumrule} is independent
of the masses of the Higgs bosons. Thus, even if the excess at 136~GeV
is not confirmed, the orange and brown regions in
Fig.~\ref{fig:modelindependentfit} will still be the favored regions
for the couplings of a second Higgs boson in any model with exactly two neutral Higgses.

It is interesting to
see how an increase in precision of the Higgs measurements would affect
the allowed $a_{h^\prime}$ -- $c_t^\prime$ parameter space.  For this,
we show in Fig.~\ref{fig:projections} 
projections assuming that all central values of measurements remain
intact, while the errors are reduced by a factor of 2 and 5. One can
see that under such assumptions, the increased  accuracy in the measurements would render the CMS data on the 136 GeV resonance incompatible with the data on the 126 GeV Higgs.

\subsection{Doubly Charged Higgses}
\label{sec:ssdl}
The primary decay modes for a doubly charged Higgs are to a pair of
same-sign $W$ bosons and to  a pair of singly charged same sign Higgs
bosons.\footnote{Assuming lepton number conservation, these are the
  only tree level, two-body decay to SM particles. Decays to $W^{\pm}
  W^{\pm} h$ or $W^{\pm} W^{\pm} Z$ are also possible, but are
  suppressed by the three-body phase space, and in some cases a mixing
  angle, and/or extra powers of $g$. At one-loop, the two-body decay
  $h^{\pm\pm} \to \ell^{\pm} \ell^{\pm}$ is possible if neutrinos are
  majorana in nature, but is suppressed by $m_{\nu} / m_{h^{++}}$
  relative to $h^{\pm\pm} \to W^{\pm} W^{\pm} \to 2\ell^{\pm} 2\nu$.}
In the analysis below we assume $\text{Br}(h^{++}\to W^+W^+)= 100\% $, since $\Gamma(h^{++}\to h^+h^+)$ is model dependent, determined by parameters in the Higgs potential.  Searches for new physics with same-sign dileptons are sensitive to $h^{++}\to W^+W^+$, and can be used to constrain the parameter space of the
doubly charged Higgs.\footnote{See Ref.~\cite{Chiang:2012dk,Kanemura:2013vxa,Chun:2013vma,delAguila:2013mia,Dermisek:2013cxa,Englert:2013wga}
  for recent studies of the bounds on and search strategies for doubly- and singly-charged scalars at the LHC.} As above, the model independent interaction is
defined as follows,
\begin{equation}
\label{eq:dch}
\mathcal{L}_{int} = g M_W b W_{\mu}^- W^{\mu -} h^{++} + \text{h.c.}.
\end{equation}

Single $h^{\pm\pm}$ production at the LHC occurs through $W$ boson fusion and in association with a $W$ boson, both of which can lead of a signature of same-sign dileptons and jets. Results of a search for this signature using the full LHC Run 1 dataset are given in Ref.~\cite{Chatrchyan:2013fea}, which expands on searches using less data~\cite{Chatrchyan:2012sa,Chatrchyan:2012paa}.\footnote{For dedicated searches for doubly-charged Higgs bosons, see~\cite{Abbiendi:2001cr,Abdallah:2002qj,Abbiendi:2003pr,Achard:2003mv,Abazov:2008ab,Aaltonen:2008ip,ATLAS:2012hi,Chatrchyan:2012ya}.} Information about event selection efficiencies is provided in~\cite{Chatrchyan:2013fea,Chatrchyan:2012sa,Chatrchyan:2012paa} such that models of NP may be constrained in an approximate way using generator-level MC studies, i.e., without performing a full detector simulation. This prescription is known to reproduce the results of the full CMS analysis to within 30\%~\cite{Chatrchyan:2012sa}.

FeynRules~\cite{Christensen:2008py} was used to implement Eq.~\eqref{eq:dch} in MadGraph 5~\cite{Alwall:2011uj} with MSTW2008 LO PDFs~\cite{Martin:2009iq}, which was used to generate events for the analysis. The kinematic requirements placed on charged leptons were $p_T > 20$ GeV (high-$p_T$ analysis) and $|\eta| < 2.4$, and on jets were $p_T > 40$ GeV and $|\eta| < 2.4$. The so-called SR5 in the high-$p_T$ analysis was the single most constraining signal region. This signal region is defined by having 2-3 jets with 0 $b$-tags, $E_T^{\text{miss}} > 120$ GeV, and $H_T \in [200,400]$ GeV. 12 events were observed in SR5 compared to an expected background of $20 \pm 7$. Using confidence interval calculator program of Ref.~\cite{Barlow:2002bk}, we place an upper limit of 6.1 non-background events at 95\% CL assuming a signal efficiency uncertainty of 13\%. The upper limit is very weakly dependent on this uncertainty such that it is still 6.1 when the signal efficiency uncertainty is taken to be 20\%. 

\begin{figure}
  \centering
\includegraphics[width=0.5\textwidth]{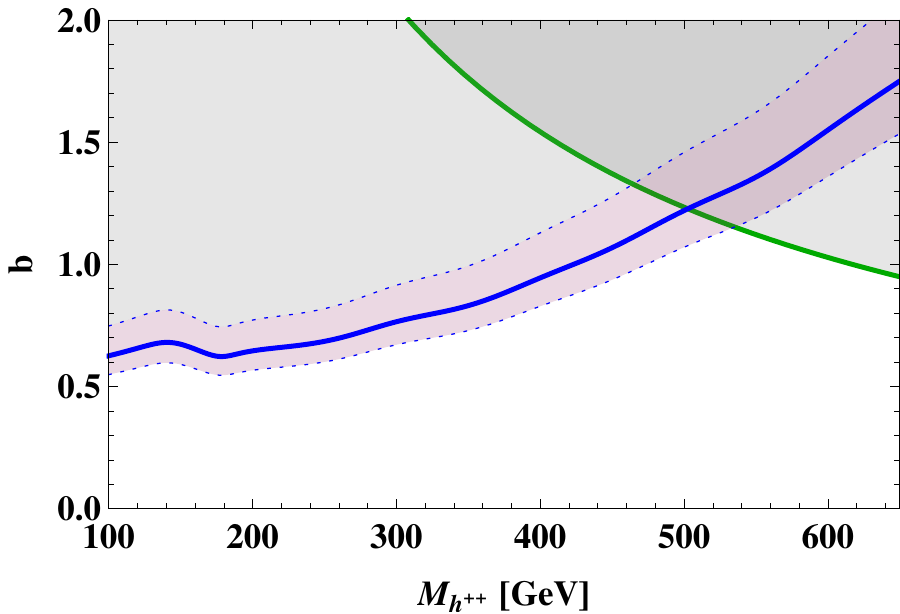}
\caption{Viable parameter space for a doubly charged Higgs based on a search for NP in events with same-sign dileptons and jets. The blue curve is the upper limit with its uncertainty given by the pink band. The green line is the perturbative unitarity bound. The parameter space shown in gray is ruled out.}
 \label{fig:dch}
\end{figure}

The results of this analysis are shown in Fig.~\ref{fig:dch}. The blue curve is the upper limit of 6.1 events with the pink band corresponding to the 30\% uncertainty in the analysis method. The parameter space above and to the left of the blue curve is ruled out by the CMS search. The green line is the perturbative unitarity bound, Eq.~\eqref{eq:massboundWW}, neglecting the 126 GeV Higgs. The parameter space above and to the right of the green curve is ruled out by perturbative unitarity.
The enhancement of the exclusion near $m_{h^{++}} \approx 2 m_W$ is due to both $W$'s in the decay chain going on-shell. Note that we did not simulate signal-background interference, which is relevant when the NP cross section approaches the SM cross section. This occurs in the pink band.  We don't claim this region is ruled out as it represents the uncertainty in the upper limit of the exclusion. We also did not perform a comprehensive study of the effect of more than one new Higgs particle as this would have required introducing several additional parameters. However, we did investigate a few benchmark points in the double-septet model with the result that the exclusion does not change significantly if the mass of the second Higgs is at least a few hundred GeV, which is expected as the Higgs cross section drops steadily with an increase in mass.

\section{Specific Models}
\label{pheno}
Consider next  various explicit realizations of electroweak symmetry breaking. For specific, perturbative models the sum rules derived by requiring that longitudinally polarized gauge boson scattering grows no faster than logarithmically with center of mass energy are automatically satisfied. But sum rules limiting the masses of various Higgs bosons are genuinely new inputs. For each of the models we study we  fit both  to the 126~GeV Higgs data and to the 126~GeV and 136~GeV Higgs data combined.  Tab.~\ref{tab:chi2} shows   
the minimum $\chi^2$ values as well as the number of degrees of freedom, $N$, in each of the models  we study in this section. The table also shows, for comparison, the results of the model independent fits of the previous section, in which  $a_h$, $c_t$, $c_b$ and $c_\tau$ are treated as independent parameters in the fit to 126 GeV Higgs data while     
  the set of independent parameters is enlarged to include $a_{h^\prime}$, $c_t^\prime$, $c_b^\prime$ and $c_\tau^\prime$  for the  fit to the combined 126 and 136 GeV data. Here $c_f$ ($c_f^\prime$) refers to the modification of neutral Higgs boson, $h$ ($h^\prime$), coupling to $\bar{f}f$ with respect to SM Higgs boson coupling.
\begin{table}[tc]
\begin{center}
\begin{tabular}{| l | c | c | c | c | }
\hline
\multicolumn{1}{|c|}{Models} &\multicolumn{2}{c|}{\phantom{\&3}126 GeV Fit\phantom{36}} &\multicolumn{2}{c|}{126 \& 136 GeV Fit} \\ \cline{2-5}
	&\phantom{a} $\chi^2/N$\phantom{a} & $N$ &\phantom{a} $\chi^2/N$\phantom{a} & $N$ \\ \hline
Model independent &0.29 &14 &0.32 &12 \\ \hline
Douplet-septet &0.31 &16 &0.71 &18\\ \hline
Georgi-Machacek &0.31 &16 &0.60 &18\\ \hline
2HDM-II &0.31 &16 &0.56 &18\\ \hline
2HDM-III & 0.28 & 14 & 0.60 & 16\\ \hline
\end{tabular}
\caption{Minimum $\chi^2$ values for various models studied. Here $N$ stands for the number of degrees of freedom.}
\label{tab:chi2}
\end{center}
\end{table}

\subsection{The Doublet-Septet Model}

The doublet-septet model contains two Higgs fields: the standard (weak) isospin-1/2 and an isospin-3 (septet) multiplets.  We have introduced earlier various aspects of this model in several different sections, so it is convenient to collect, review and expand on them here.  As noted above, the septet contains a doubly charged Higgs, whose interactions with the vector bosons can be written as in \ref{sec:couplings},
\begin{equation} \label{eq:dchLag}
\mathcal{L}_{int} \supset\sqrt{15} \frac{M_W^2}{v_{\text{EW}}}\cos\beta\left(W^{-}_{\mu} W^{-\mu} h^{++} + W^{+}_{\mu} W^{+\mu} (h^{++})^{\star}\right)~,
\end{equation}
where the mixing angle, $\beta$, is defined as $\tan\beta = v_1/(4 v_2)$.  
Here $v_1$ ($v_2$) denotes the VEV of the doublet (septet).  Note that $M_W^2 \cos\beta / v_{\text{EW}} = g^2 v_2$,
or $4 v_2 = v_{\text{EW}} \cos \beta$ and $v_1 = v_{\text{EW}} \sin
\beta$.  The interactions of the neutral Higgses are given by 
\begin{equation}
\mathcal{L}_{int} \supset \frac{2}{v_{\text{EW}}} \left(M_W^2 W^{+}_{\mu} W^{-\mu} + \frac{1}{2} M_Z^2  Z_{\mu} Z^{\mu}\right)\left((s_{\beta} c_{\alpha} - 4 c_{\beta} s_{\alpha}) h + (s_{\beta} s_{\alpha} + 4 c_{\beta} c_{\alpha}) h'\right),
\end{equation}
where $h,h'$ are the neutral Higgs states in the mass basis, related
to those in the weak basis, $h^0_2$ and $h^0_7$,  by
\begin{equation}
\begin{pmatrix}
h^0_2 \\
h^0_7
\end{pmatrix}
=
\begin{pmatrix}
c_{\alpha} & s_{\alpha} \\
-s_{\alpha} & c_{\alpha}
\end{pmatrix}
\begin{pmatrix}
h \\
h'
\end{pmatrix},
\end{equation}
with $s_{\alpha}$, $c_{\alpha}$ standing for the sine and cosine of $\alpha$, the mixing angle between the weak mass bases. We identify $h$ with the 126 GeV resonance, recently discovered at the LHC~\cite{Aad:2012tfa,Chatrchyan:2012ufa}. The couplings of the neutral Higgses to
fermions are given by
\begin{equation}
\mathcal{L} \supset \left(\frac{\cos\alpha}{\sin\beta} h + \frac{\sin\alpha}{\sin\beta} h^{\prime} \right)\sum_i \frac{m_{f,i}}{v_{\text{EW}}}\bar{f}_i f_i,
\end{equation}
where $m_{f,i}$ denotes the mass of the $i^{\text{th}}$ fermion. The doubly charged Higgs obviously cannot couple to fermions at the
renormalizable level. The above couplings give  the
parameters $a_i$ and $b$ displayed in \eqref{eq:physcouplings},
while the couplings to fermions are parametrized by
\begin{equation*}
c_f = \cos\alpha / \sin\beta, \quad c_f' = \sin\alpha / \sin\beta.
\end{equation*}

\begin{figure}
  \centering
 \includegraphics[width=0.5\textwidth]{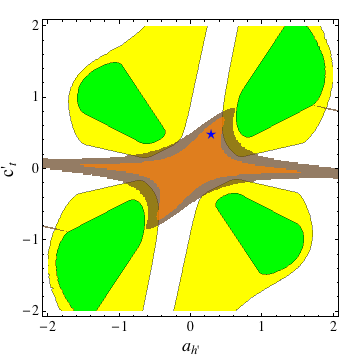}
 \caption{Allowed region of $a_{h^\prime}$-$c_t^\prime$ for the
   doublet-septet model from a fit to the 126~GeV Higgs data. The best
   fit value is indicated by a blue star, and the orange and brown
   regions give the 68\% and 95\% CL regions, respectively.  The
   underlying green/yellow regions are from the model independent fit to the 136~GeV
   data  in Fig.~\ref{fig:136}. }
 \label{fig:septet}
\end{figure}

One can infer the bounds on the extra neutral resonance from the available data on the 126 GeV Higgs in much the same way as for the case of the model-independent fit outlined above. The fit to the 126 GeV data, combined with the condition of perturbative unitarity in the $WW\to t\bar t$ channel significantly reduces the allowed parameter space for the $136$~GeV Higgs couplings. The 68\% and 95\% CL regions in the $a_{h^\prime} - c_t^\prime$ plane (defined in the same way as for the model-independent fit) allowed by the 126 GeV data are shown in orange and brown in Fig.~\ref{fig:septet}.  The best fit
value is marked by a star; the corresponding (minimum) has
$\chi^2_{DS} = 4.96$ for 16 degrees of freedom. The figure shows limited overlap between the model independent  fit
to the 136~GeV data (in green/yellow) and the model specific fit to the 126 GeV data (in
orange/brown). Alternatively,  a  global fit to both 126 and 136
  GeV data gives a higher  minimum, with   $\chi^2_{DS}$ at  12.71 for 18 degrees of freedom. 

Another bound on a second Higgs can be obtained from the
ATLAS and CMS searches for SM-like Higgs bosons in the $WW$ and $ZZ$
channels~\cite{ATLAS-CONF-2013-013,ATLAS-CONF-2013-067,CMS-PAS-HIG-12-024,CMS-PAS-HIG-13-002,CMS-PAS-HIG-13-003,CMS-PAS-HIG-13-008,CMS-PAS-HIG-13-014}. Taken
at face value, our combination of these bounds in
Appendix~\ref{sec:app} rules out Higgs bosons that couple to $WW$ and
$ZZ$ with SM strength at 95\% CL for $m_{h^{\prime}} = 128-1000$
GeV. However, there is no reason to expect 
neutral Higgs particles should couple to EW gauge bosons with SM
strength in multi-Higgs models. The couplings must satisfy the
model-dependent and model-independent sum rules,
Eqs.~\eqref{eq:sum0} and~\eqref{sumrule4}, respectively, but
both of these allow for a  range of coupling strengths. Instead, for a given set of parameters in a model, one must compare the predicted signal strength against the curve in Fig.~\ref{fig:comb} to determine the range(s) for which $m_{h^{\prime}}$ is ruled out. The signal
strength of $h'$ in the $WW+ZZ$ channel is given by\footnote{Here we ignore the small tree-level branching ratios into light fermions and loop-induced branching ratios into $\gamma\gamma$ and $Z\gamma$.}
\begin{equation} 
\label{eq:muH}
\mu(h' \to WW + ZZ) = \frac{c_{f}^{\prime2} \sigma_{ggF + t\bar{t}h'} + 
                           a_{h'}^2 \sigma_{VBF + Vh'}}{\sigma_{ggF + t\bar{t}h'} + 
                           \sigma_{VBF + Vh'}} \frac{ a_{h'}^2}{c_{f}^{\prime2} Br_{f\bar{f}} + a_{h'}^2
 Br_{VV} + c_{f}^{\prime2} Br_{gg}},
\end{equation}
where $\sigma$ ($Br$) is the expected SM cross-section (branching ratio) for the Higgs boson with mass $m_{h^\prime}$, with $f = \tau, c, b, t$,  $V = W, Z$.  All cross sections and branching ratios
in~\eqref{eq:muH} are taken from Ref.~\cite{Dittmaier:2011ti}. Here we implicitly assume $h^\prime$ has the same decay channels available  as $h$, but if heavy enough new channels may become available, e.g.,  it could decay to a pair of scalars and/or a pseudoscalar and a neutral vector boson ($AZ$), and these channels may have significant branching fractions.  See Ref.~\cite{Grinstein:2013npa} for an analysis of these decay modes in the Type-II two Higgs doublet model.

\begin{figure}
\vspace{-100pt} 
  \centering
  \subfloat{\label{fig:cp135}\includegraphics[width=0.4\textwidth]{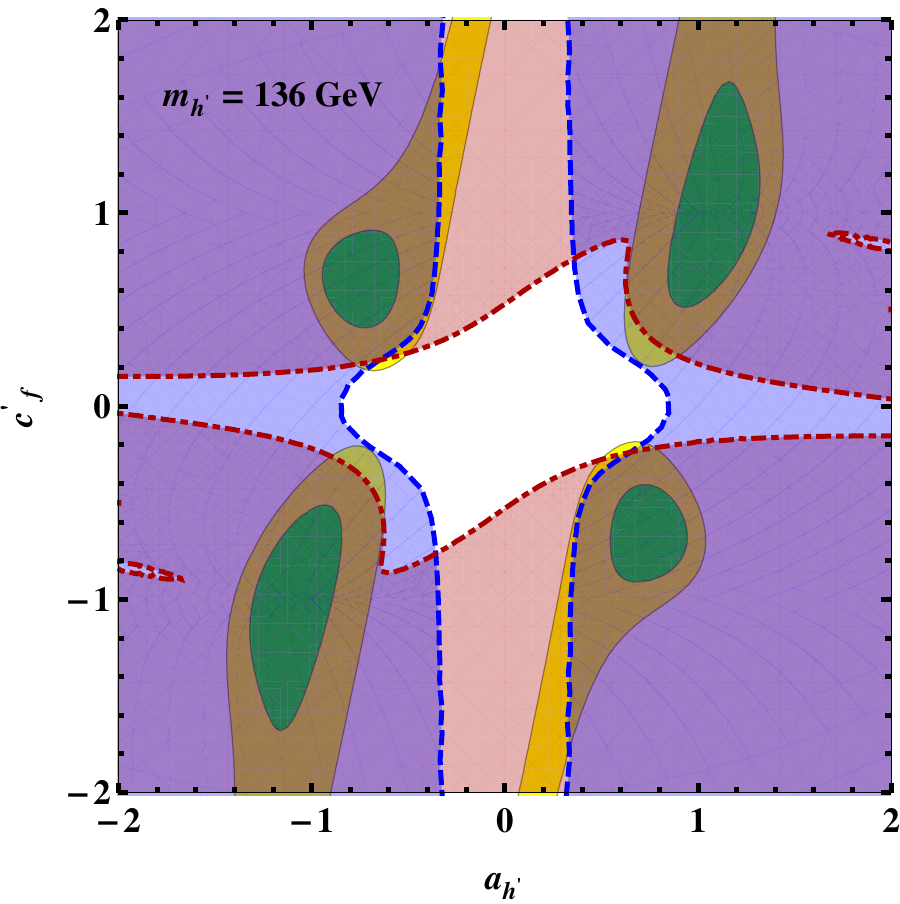}} \hspace{5pt}
  \subfloat{\label{fig:cp250}\includegraphics[width=0.4\textwidth]{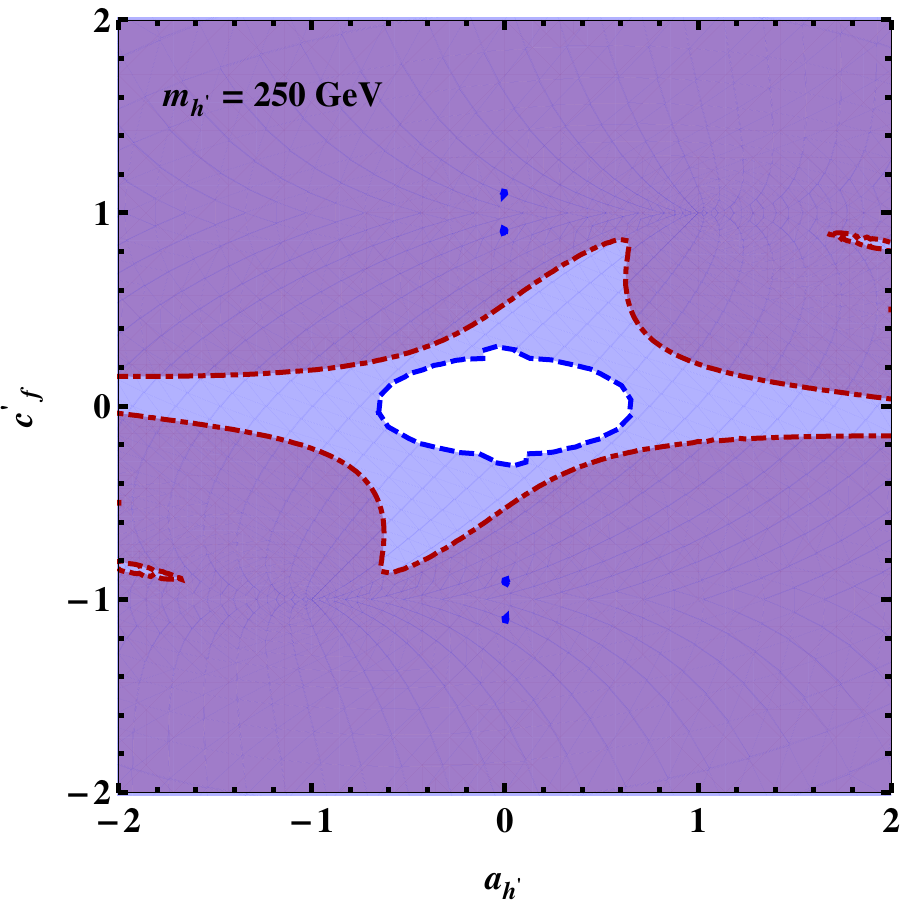}} \\
  \subfloat{\label{fig:cp500}\includegraphics[width=0.4\textwidth]{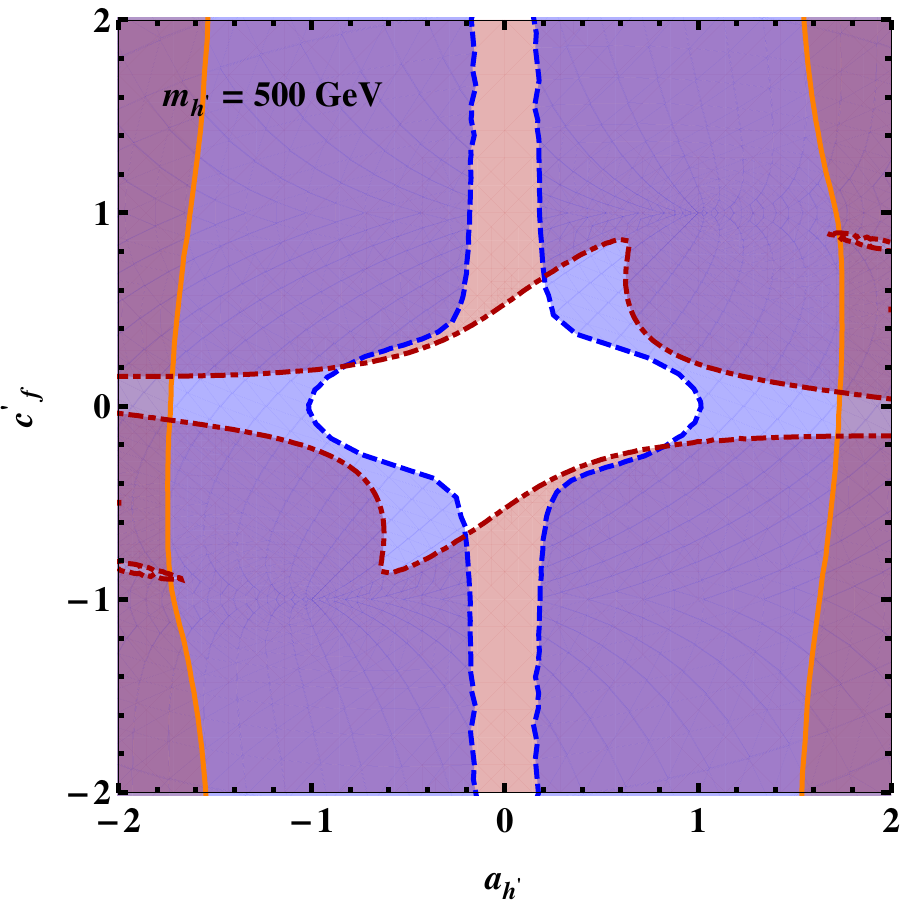}} \hspace{5pt}
  \subfloat{\label{fig:cp750}\includegraphics[width=0.4\textwidth]{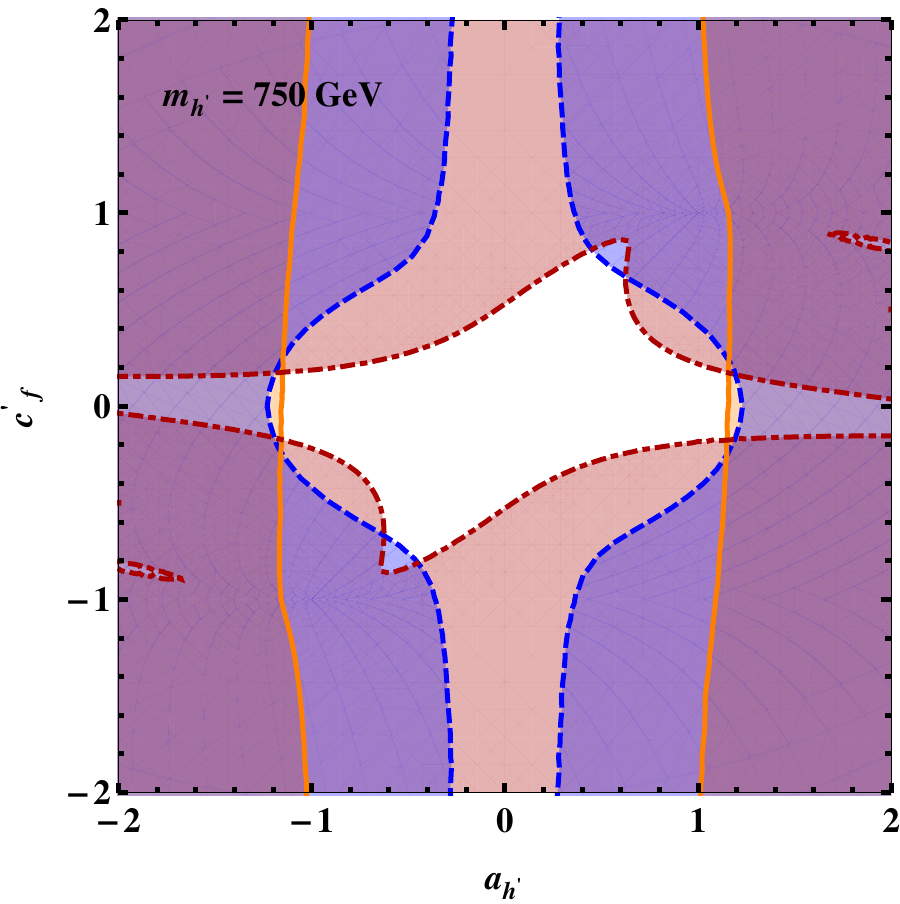}} \\
  \subfloat{\label{fig:cp1000}\includegraphics[width=0.4\textwidth]{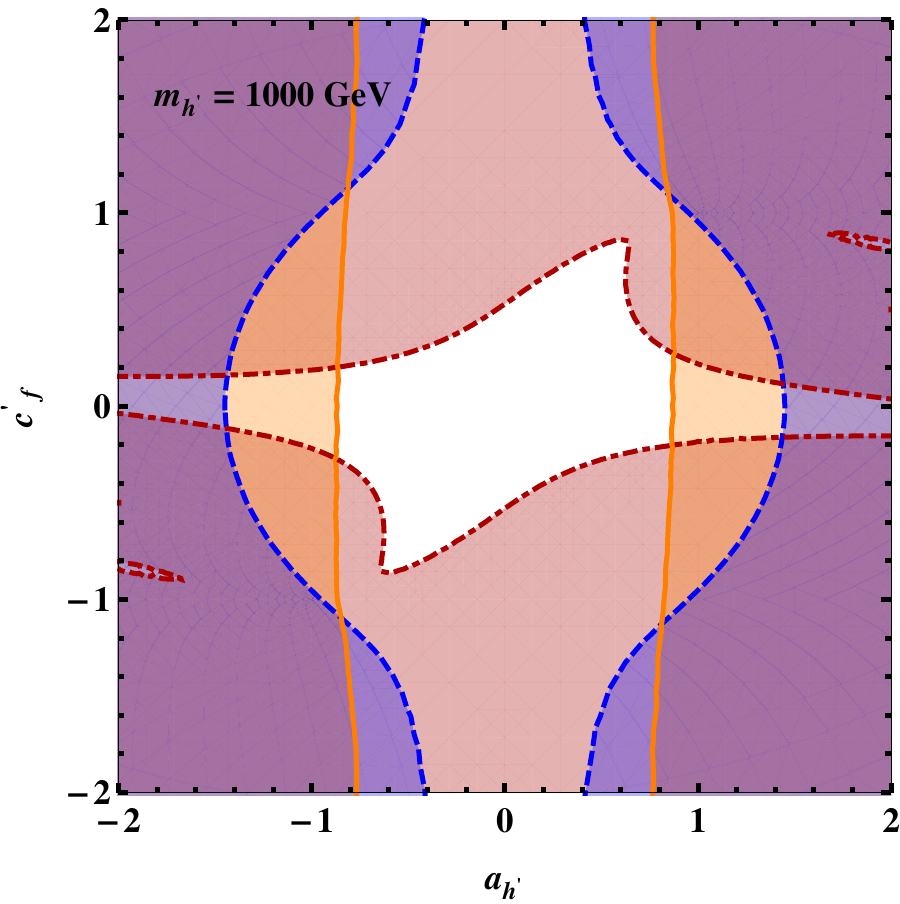}} \hspace{5pt}
  \subfloat{\label{fig:cp1500}\includegraphics[width=0.4\textwidth]{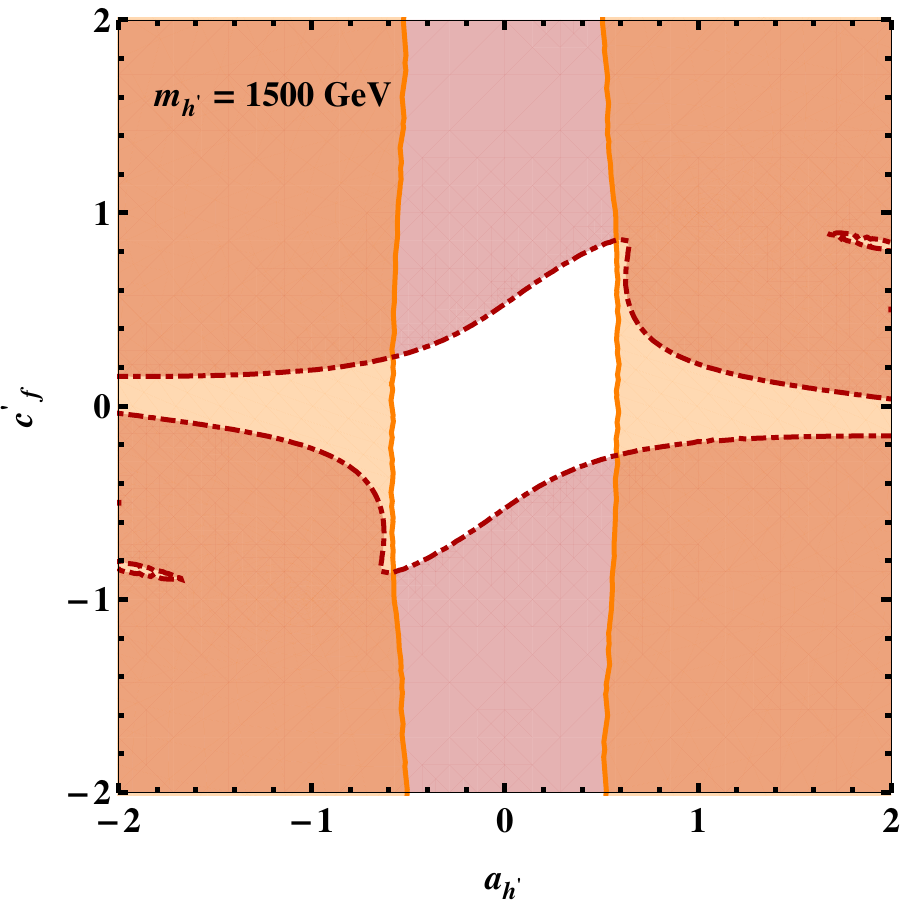}} 
  \caption{Parameter space in the doublet-septet model that is ruled out at the 95\% CL due to searches for heavy Higgses (blue), 126 GeV Higgs data (red) and perturbative unitarity (orange) for various values of $m_{h^{\prime}}$. For the $m_{h^{\prime}}=136~\text{GeV}$ case, the figure is superimposed on Fig.~\ref{fig:136cwm} that shows the 68\% and 95\% CL model-independent regions compatible with the excess seen by CMS, displaying little overlap with the allowed (white) region from this analysis. }
  \label{fig:cpmasses}
\end{figure}

For given values of $a_{h^{\prime}}$ and $c^\prime_{f}$ (or $\alpha$ and $\beta$), $\mu(h^{\prime} \to WW + ZZ) $ as a
function of $m_{h^\prime}$ can be compared against the data in Fig.~\ref{fig:comb}.  The parameter space that is ruled out by searches for additional Higgs bosons is given by the blue regions in Fig.~\ref{fig:cpmasses} for 6 different values of $m_{h^{\prime}}$. In addition, we show in red the complement of the 95\% CL region of  parameter space from the fit to the 126 GeV Higgs data (the complement to the orange and brown region in Fig.~\ref{fig:septet}). Finally, the orange region is ruled out by perturbative unitarity. It is essentially the bound from Fig.~\ref{fig:uni1} plotted as a function of $a_{h^{\prime}}$ and $c^{\prime}_{f}$ rather than  $\alpha$ and $\beta$. The bounds in Fig.~\ref{fig:136cwm} have consequences for the interpretation of the excess at 136 GeV in~\cite{CMS-PAS-HIG-13-016} as a second neutral Higgs boson. In the upper-left panel of Fig.~\ref{fig:cpmasses} we also plotted the allowed region for $a_{h^{\prime}}$ and $c^{\prime}_f$ from  Fig.~\ref{fig:136cwmb}. We see that  for the 136 GeV Higgs in the doublet-septet model  there is very limited overlap between the 95\% CL-allowed region from the exclusion limits from searches for heavy Higgs bosons in the $WW$ and $ZZ$ channels and the 95\% CL-allowed region from data on the 126 GeV Higgs.

\subsection{Other Models}
In this subsection we compare the result from the model independent fit to explicit models with extended Higgs sectors.
In all the models we consider  electroweak symmetry breaking is perturbative.  

\subsubsection{Georgi-Machacek Model}

\begin{figure}
  \centering
 \includegraphics[width=0.5\textwidth]{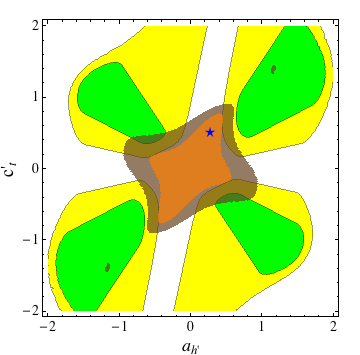}
\caption{Region of the $a_{h^\prime}$-$c_t^\prime$ parameter space consistent with the fit to the GM model. The orange and brown regions  are compatible at 68\% and 95\% CL, respectively and the best fit value is indicated  by a blue star. The
   underlying green/yellow regions are from the model independent fit to the 136~GeV
   data  in Fig.~\ref{fig:136}. The darker blotches contained within the green region are overlaps of brown and green. }
 \label{fig:GM}
\end{figure}

The Georgi-Machacek(GM) model consists of one electroweak doublet of hypercharge~1/2 and two electroweak triplets with hypercharge 0 and 1. 
For a recent phenomenological study of the model see Ref.~\cite{Chiang:2012cn} and references therein.\footnote{We follow the conventions of Ref.~\cite{Chiang:2012cn} with the exception that the mixing angles $\alpha$ and $\beta$ below are obtained through the replacement in \cite{Chiang:2012cn}: $\alpha\to-\alpha$, and $\beta\to\frac{\pi}{2}-\beta$.}
The electroweak VEV is given by $v_{\text{EW}}^2 = v_\phi^2 + 8v_\Delta^2$ with $\tan\beta = v_\phi/2\sqrt{2}v_\Delta$, where $v_\phi$ and $v_\Delta$ are the VEVs of the doublet and the two triplets respectively.
There are three CP even neutral scalar fields in the GM model, $\phi_r$, $\tilde{H}_1^0$ and $H_5^0$.  These fields are related to the physical states by
\begin{equation}
	\begin{pmatrix}\phi_r\\ \tilde{H}_1^0\\ H_5^0\end{pmatrix} = 
	\begin{pmatrix}c_\alpha & s_\alpha & 0\\ -s_\alpha & c_\alpha & 0\\ 0 &0 &1\end{pmatrix}
	\begin{pmatrix}h\\ h^\prime\\ H_5^0\end{pmatrix},
\end{equation}
where $c_\alpha = \cos\alpha$ and $s_\alpha = \sin\alpha$. The couplings to the vector boson and fermion pairs for the two Higgses are given as follows
\beq
a_h=\frac{1}{3}\(3 c_\alpha s_\beta-2\sqrt{6} s_\alpha c_\beta\),~ c_f=\frac{c_\alpha}{s_\beta},~ a_{h'}=\frac{1}{3}\(3 s_\alpha s_\beta+2\sqrt{6} c_\alpha c_\beta\),~ c'_f=\frac{s_\alpha}{s_\beta}~.\nn
\eeq
By performing a fit to the 126 GeV data, we obtain a minimum $\chi^2_{GM} = 4.96$ for 16 degrees of freedom.
Within the context of the GM model, the couplings $c_{h^\prime}$ and $c_t^\prime$ consistent with the model-independent fit are shown in Fig.~\ref{fig:GM} where again the best fit value is marked by a star.  
Alternatively,  a  global fit to both 126 and 136
  GeV data gives a higher  minimum   $\chi^2_{GM}$ at 10.84 for 18 degrees of freedom.

\subsubsection{Two-Higgs-Doublet Model}

\begin{figure}
  \centering
 \includegraphics[width=0.45\textwidth]{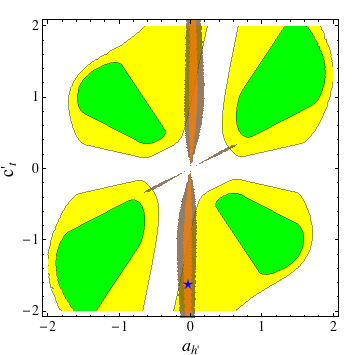}
 \includegraphics[width=0.45\textwidth]{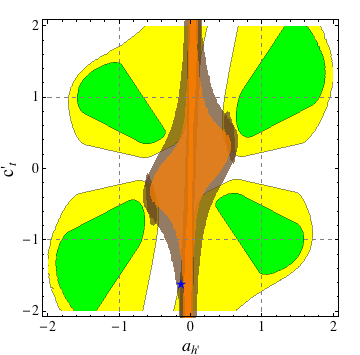}
\caption{Region of the $a_{h^\prime}$-$c_t^\prime$ parameter space consistent with the fit to the  Type-II (left) and Type-III (right) 2HDM.  The orange and brown regions  are compatible at 68\% and 95\% CL, respectively. The best fit value for the Type-III model  is indicated  by a blue star, while it  is at ($a_{h^\prime},\,c_t^\prime$) =
   (-0.02,-3.14)  for the  Type-II 2HDM. The
   underlying green/yellow regions are from the model independent fit to the 136~GeV
   data  in Fig.~\ref{fig:136}.}
 \label{fig:2hdm2}
\end{figure}

There are many variations of the two-Higgs-doublet models, see Ref~\cite{Branco:2011iw} for a recent review.  Here we will focus on two explicit realizations, the so called type-II (2HDM-II) and type-III (2HDM-III).  In 2HDM-II a discrete $Z_2$ symmetry is imposed to forbid tree-level flavor changing neutral currents (FCNC).  In this case the neutral scalar couplings to vectors and fermions are determined by two angles, $\alpha$ and $\beta$. The angle $\alpha$ is the mixing angle for the two CP-even neutral scalars, while $\tan \beta$ is given by the ratio of the two VEVs.
The coupling modifiers for the two Higgses are given in this case by the following expressions
\beq
a_h=\sin(\alpha+\beta),~c_t=c_c=\frac{c_\alpha}{s_\beta},~ c_b=c_\tau=\frac{s_\alpha}{c_\beta},~a_{h'}=\cos(\alpha+\beta),~c'_t=c'_c=\frac{s_\alpha}{s_\beta},~ c'_b=c'_\tau=\frac{c_\alpha}{c_\beta}.\nn
\eeq
 
By performing a fit to the 126 GeV data, we obtain a minimum $\chi^2_{2HDM-II} = 4.91$ for 16 degrees of freedom.
Within the context of the 2HDM-II, the couplings $a_{h^\prime}$ and $c_t^\prime$ consistent with the model-independent fit are shown in Fig.~\ref{fig:2hdm2}.  
Alternatively,  a  global fit to both 126 and 136
GeV data gives a higher  minimum   $\chi^2_{2HDM-II}$ at 10.13 for 18 degrees of freedom. 

In 2HDM-III where the discrete $Z_2$ symmetry is not imposed, there in general are flavor changing neutral currents.  However, we will ignore FCNC and proceed to study the neutral scalar couplings to vectors and fermions in this model.  Unlike in the 2HDM-II, here the couplings to the third generation fermions, relevant for Higgs phenomenology, are characterized by three additional parameters $c_t$, $c_b$ and $c_\tau$ (the couplings to the vector bosons are the same as in 2HDM-II). Fitting to the 126 GeV data, we obtain a minimum $\chi^2_{2HDM-III} = 3.93$ for 14 degrees of freedom, while fitting to both 126 and 136
GeV data gives a higher  minimum   $\chi^2_{2HDM-III}$ at 9.61 for 18 degrees of freedom.

\section{Consistency with dispersion relations}
\label{disper-rels}

Dispersion relations have been used to constrain models with zero
\cite{Distler:2006if} and one \cite{Low:2009di, Falkowski:2012vh,Urbano:2013aoa} Higgs
particles. It is straightforward to generalize those to the case of
multi-Higgs models. For self-completeness, we give a detailed derivation of the dispersion relations of Refs. \cite{Low:2009di, Falkowski:2012vh,Urbano:2013aoa} and their multi-Higgs generalizations in appendix \ref{sec:appb}. 

One such dispersion relation, that is of particular interest to us, is written as follows
\begin{align}
\label{eq:disp1}
&\frac{d\CA}{ds}\bigg|_{s=0}=\frac{2w^0+w^1-w^2}{\pi}\int_{0}^\infty \frac{ds}{s} \left[
\sigma_{+-}-\sigma_{++}\right], 
\end{align}
where $\sigma_{+-}$ denotes the total cross section for a
longitudinally polarized $W^+W^-$ scattering, and likewise for
$\sigma_{++}$.  Alternatively, using the equivalence theorem one may
calculate $\sigma_{+-}$ and $\sigma_{++}$ as the cross section for
$\pi^+\pi^-\to \textit{anything}$ and $\pi^+\pi^+\to
\textit{anything}$, in a nonlinear sigma model for electroweak
symmetry breaking. Below we refer to this as ``pion scattering,'' keeping in mind
that we are really describing longitudinally polarized vector boson
scattering.  The amplitude $\CA$ is expanded in terms of amplitudes of
definite isospin (isospin-$0,1,2$ respectively), $\CA=\sum_I w^I
T^I$, with coefficients $w^I$. The integral in the once subtracted
dispersion relation \eqref{eq:disp1} is not generally convergent. Barring a cancellation in the difference of cross sections, a cross section saturating the Froissart bound would give a divergent integral. Therefore, in order to be able to use \eqref{eq:disp1} \emph{e.g.} for strongly coupled theories, one has to assume that the UV cross-sections exhibit milder high-energy behavior than allowed by the Froissart bound.

On the other hand, for perturbative models with definite UV field content like the ones considered above, the cross sections do give convergent integrals. However, in those cases both sides of the dispersion relation can be calculated perturbatively to check that it is identically satisfied, once unitarity constraints are imposed. 

One can readily study the implications of the dispersion relation \eqref{eq:disp1}  and their consistency with the relations obtained from perturbative unitarity above. Let us for definiteness concentrate on $\pi^+\pi^-$ scattering. Using eqs. \eqref{eq:chargeamplitudes} and \eqref{eq:CAofT} of appendix \ref{sec:appb}, the decomposition of the corresponding amplitude into the isospin eigenbasis can be written as follows, $\CA_{\pi^+\pi^-}=\sum_I w^I T^I$, with $w^{0,1,2}=(1/3,~1/2,~1/6)$, so that $2 w^0+w^1-w^2=1$. The once subtracted dispersion relation for charged pion scattering thus implies
\beq
\label{eq:finaldr1}
\frac{d\CA_{\pi^+\pi^-}}{ds}\bigg|_{s=0}=\frac{1}{\pi}\int_{0}^\infty
\frac{ds}{s} \left[ \sigma_{+-}-\sigma_{++}\right]~.  \eeq Assuming,
as required by unitarity at order $\hbar^0$, that all contributions to
the amplitude that grow with $s$ cancel, one readily obtains the following
tree-level\footnote{By ``tree-level'' we refer to amplitudes that are
  of order $1/v^2$ in the derivative expansion.} expression for the
finite piece of the forward scattering amplitude 
\beq
\label{eq:finaldr}
\CA_{\pi^+\pi^-}=\frac{s}{v^2}\(\sum_i \frac{a_i^2 m_{h_i}^2}{m_{h_i}^2-s}-\sum_r\frac{4b_r^2 M_r^{++2}}{M_r^{++2}+s}\)~.
\eeq
Here the sums are performed over various neutral and doubly charged Higgs bosons. Straightforward differentiation then yields
\beq
\frac{d\CA_{\pi^+\pi^-}}{ds}\bigg|_{s=0}=\frac{\sum_i a^2_i - 4\sum_r b_r^2}{v^2}~.
\eeq

\begin{figure}
  \centering
 \includegraphics[width=0.65\textwidth]{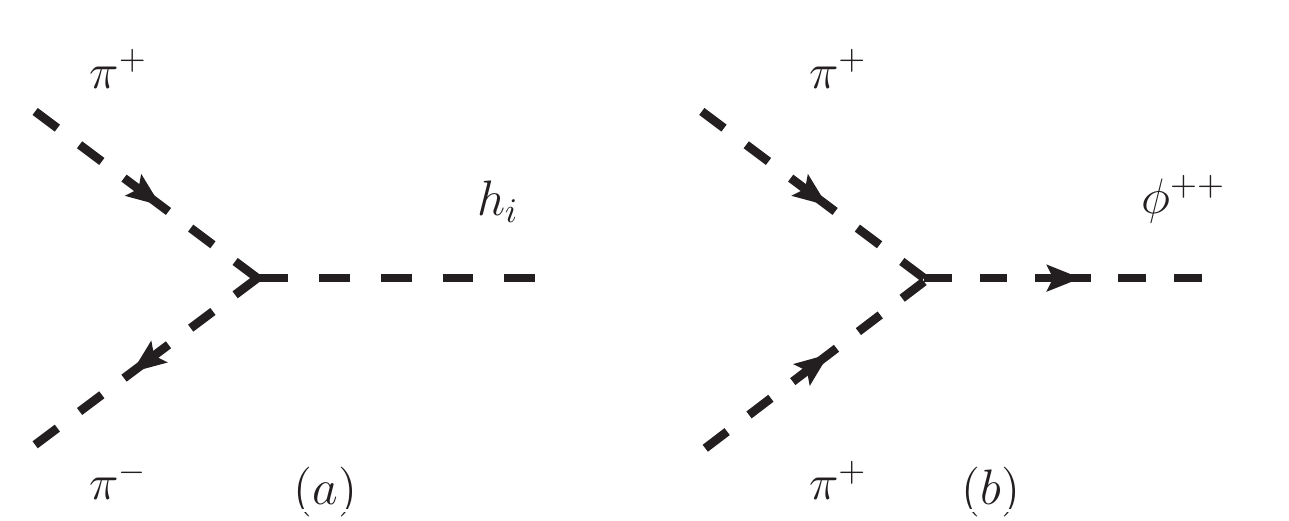}
\caption{Diagrams contributing to (a) $\sigma_{+-}$  and  (b)  $\sigma_{++}$ in \eqref{eq:finaldr1}.}
 \label{fig:sigmas}
\end{figure}

Furthermore, at order $1/v^2$, there is a single diagram contributing to each of the terms on the right hand side of \eqref{eq:finaldr}; the tree-level annihilation $\pi^+\pi^-\to h_i$ contributes to $\sigma_{+-}$, while $\sigma_{++}$ receives its only contribution from $\pi^+\pi^+\to \phi_r^{++}$.
These cross sections are given as follows
\beq
\label{eq:sigmas}
\sigma_{+-}=\sum_i \frac{\pi a_i^2 s }{2 v^2 m^2_{h_i}}\delta\(\sqrt{s}-m_{h_i}\), \quad  \sigma_{++}=\sum_r\frac{4 \pi b_r^2 s }{2 v^2 M_r^{++2}}\delta\(\sqrt{s}-M^{++}\)~,
\eeq
where the delta functions remain after integration over the single particle phase space. Using these expressions, it is straightforward to show that 
\beq
\frac{1}{\pi}\int_{0}^\infty \frac{ds}{s} \left[
\sigma_{+-}-\sigma_{++}\right]=\frac{\sum_i a^2_i - 4 \sum_rb_r^2}{v^2}=\frac{d\CA_{\pi^+\pi^-}}{ds}\bigg|_{s=0}~.
\eeq
As advertised, the dispersion relation is identically satisfied for theories that respect unitarity at order $\hbar^0$. The only input that we used from unitarity in the above analysis is that the amplitudes do not grow at large center of mass energy, so that the integral of the cross sections in the once subtracted dispersion relation is convergent and well-defined. In fact, at the (tree) level we are interested in here, the singularity structure on the complex $s$ plane is drastically simplified. Indeed, in the limit $\hbar\to 0$ there are no cuts on the real axis,  only singularities corresponding to on-shell poles  from heavy particle exchange contribute. The imaginary part of amplitudes corresponding to exchange of a particle of mass $M$ is proportional to
\beq
\text{Im} \frac{1}{p^2-M^2+i\epsilon}=-\pi \delta(p^2-M^2)~,
\eeq
and only has support on the heavy particle pole. This is also evident from \eqref{eq:sigmas}, given that the cross section is related to the imaginary part of the forward scattering amplitude through the optical theorem. The discontinuity of the amplitude across these poles gives the only contribution to the right hand side of \eqref{eq:finaldr1} at tree level. It is straightforward to extend this analysis beyond tree level, but the main result, that the dispersion relation is identically satisfied for unitary theories that are weakly coupled over the full range of energy scales, is unchanged. Moreover, the dispersion relations do not imply the perturbative unitarity bounds on Higgs masses obtained in Sec.~\ref{sec:unitarity}.

\section{Discussions and Conclusions}
\label{concl}

Charged scalars play a crucial role in unitarizing the vector-vector scattering amplitudes in a model with a weakly coupled Higgs sector.  
In the case of $W^+_LW^-_L\to W^+_LW^-_L$ scattering, the full amplitude does not grow with $s$ due to a cancellation between $s$- and $t$- channel contributions from neutral Higgses and $u$- channel contributions from doubly-charged Higgses. Similarly, the combined contribution of neutral and singly-charged Higgs bosons ensure that the amplitude for $Z_LZ_L\to W^+_LW^-_L$ does not grow with $s$. Not surprisingly, this cancellation holds for an arbitrary number of Higgs multiplets with arbitrary representations under $SU(2)_L \times U(1)_Y$.

Unitarity also places constraints on the spectrum of the Higgses and their couplings to fermions.    
 The constraint on the spectrum depends strongly on the number and the representation of the multiplets involved in electroweak symmetry breaking. 
We studied this constraint in the doublet-septet model in section~\ref{subsec:massbound}.  
The constraint on the couplings of neutral Higgses to fermions arises from unitarity requirement in $W^+_LW^-_L\to f\bar{f}$ channel.
Unlike the bound on the spectrum, where detailed knowledge of Higgs multiplets is needed, the constraint on the couplings to fermions only depend on the knowledge of the number of neutral Higgses present in the model and not on the masses of the Higgs bosons.  

Now we turn to discuss the interpretation of the 136 GeV excess observed in CMS diphoton signal as a second neutral Higgs.
Since this excess has been only observed in the diphoton channel, we focus on studying the coupling of the putative second neutral Higgs to vector boson and top-quark (characterized by $a_{h^\prime}$ and $c^\prime_t$ respectively).
We can be predictive on the couplings $a_{h^\prime}$ and $c^\prime_t$ if we  assume that there are only two CP-even neutral Higgses in the spectrum.
In this case unitarity in $W^+_LW^-_L\to t\bar{t}$ channel requires $a_{h^\prime} c^\prime_t = 1 - a_h c_t$.  
By deducing the coupling $a_h$ and $c_t$ from the 126 GeV Higgs data, we obtained the model-independent prediction of the coupling $a_{h^\prime}$ and $c^\prime_t$ shown in figure~\ref{fig:modelindependentfit}.
We found  there that the interpretation of the 136 GeV excess is in some tension with the 126 GeV Higgs data.
Our projection of the situation assuming improved experimental precision in both the 126 and 136 GeV data show that if the 136 GeV excess persists, it cannot be explained by a second neutral Higgs in a model with only two neutral CP-even Higgses.

We can in fact improve on the prediction of the couplings of the
second Higgs in Fig.~\ref{fig:modelindependentfit} by further
specifying the origin of the second Higgs particle.
Figs.~\ref{fig:septet}, \ref{fig:GM} and~\ref{fig:2hdm2} show the
resulting fit in sample specific models, showing in all cases a
reduction in the allowed region of parameter space. However, the
allowed regions are qualitatively different for the various models we
investigated, with obvious implications for the production mechanism
in collider experiments. The most striking of these is for the 2HDM
Type-II model, Fig.~\ref{fig:2hdm2} (left pannel) , for which vector
boson fusion is largely inoperative while the rate for gluon fusion
could differ vastly from that of a SM Higgs.

\appendix

\section{Physical Higgses Couplings}
\label{sec:app0}
In this appendix we collect the couplings of the physical Higges.  
We denote the physical CP-even neutral Higgses by $h_i$, the singly charged Higgses by $\phi^+_j$ and the doubly charged by $\phi^{++}_r$. 
The couplings can be parametrized by
\begin{equation}
\label{eq:physicalcouplings}
\begin{split}
	\mathcal{L}_{phys} &\supset gM_WW^{+\mu}W^-_\mu\sum_ia_ih_i + \frac{gM_W}{2}Z^\mu Z_\nu\sum_i d_i h_i - \frac{m_f}{v_{\text{EW}}}\sum_ic_i\bar{f}fh_i\\
	&\quad + gM_W \left(Z^\mu W^+_\mu\sum_j f^{\phantom{+}}_j\phi^+_j+\hc\right) + gM_W \left(W^{+\mu}W^+_\mu \sum_rb^{\phantom{+}}_r\phi^{++}_r+\hc\right).\\
\end{split}
\end{equation}
The value of the couplings $a$, $b$, ..., $f$ depends on the detail of the Higgs sector--- the number of electroweak multiplets and the size of each multiplet.  

Here we recall the couplings of the Higgses in a generic representation of the electroweak gauge group from Sec.~\ref{sec:unitarity}.
This result can be easily extended to a case with an arbitrary number of electroweak multiplets.
For definiteness, we take the Higgs field, $\Phi$, to transform in a $(2n+1)$-dimensional representation of $SU(2)_L$ with hypercharge $-m$, $-n\le m\le n$.
To preserve electric charge the VEV of the multiplet must be in the $T^3=m$ component, $\langle\phi_m\rangle=v/\sqrt{2}$. 
Thus $\phi_{m\pm2}$ corresponds to the doubly charged component and $\phi_{m\pm1}$ is the singly charged component.
Their interactions are
\begin{equation}
\begin{split}
	\mathcal{L}&\supset \frac{g^2v}{\sqrt2}\eta^{\mu\nu}\left[A\left(W^+_\mu W^+_\nu\phi_{m-2}+W^-_\mu W^-_\nu\phi_{m-2}^*\right)
	+B\left(W^+_\mu W^+_\nu\phi_{m+2}^*+W^-_\mu W^-_\nu\phi_{m+2}\right)\right]\\
	&+\frac12\eta^{\mu\nu}\left[g^2W^+_\mu W^-_\nu\left(n(n+1)-m^2\right)
         +(g^2+g^{\prime2})Z_\mu Z_\nu m^2\right](v+\phi_m)^2\\
  &\qquad + \frac{v\cos\theta_W}{\sqrt{2}}~ \bigg [ Z^\mu W^{+}_\mu
  \(D\phi_{m-1}+E\phi^*_{m+1}\) + h.c.\bigg ],
\end{split}
\end{equation}
where $g$ $(g')$ is the $SU(2)$ ($U(1)$) couplings and
\begin{equation}
\begin{aligned}
	A &=\frac{1}{2}\sqrt{\left[n(n+1)-(m-2)(m-1)\right]\left[n(n+1)-m(m-1)\right]},\\
	B &=\frac{1}{2}\sqrt{\left[n(n+1)-m(m+1)\right]\left[n(n+1)-(m+1)(m+2)\right]},\\
	D &=F \big[2 (g^2+g'^2)m-g^2 \big ],\\
	E &=G \big[2 (g^2+g'^2)m+g^2 \big ],\\
	F &=\sqrt{\tfrac12\big(n(n+1)-m(m-1)\big)},\\
	G &=\sqrt{\tfrac12\big(n(n+1)-m(m+1)\big)}.
\end{aligned}
\end{equation}
We can relate the above parameters to the couplings of the physical Higgses, as in Eq.~\eqref{eq:physicalcouplings}.
The couplings of the physical Higgses are given by 
\begin{equation}
\begin{aligned}
	gM_W a &= g^2v\left[n(n+1)-m^2\right],\\
	gM_W b_{1(2)} &= \frac{g^2v}{\sqrt{2}}A\,(B),\\
	gM_W d &= 2(g^2+g^{\prime2})v\,m^2,\\
	(gM_W f)^2 &=\(\frac{v\cos\theta_W}{\sqrt{2}}\,
\frac{4m\,F \,G(g^2+g^{\prime2})}{\sqrt{F^2+G^2}}\)^2.
\end{aligned}
\end{equation}
Note that one linear combination of $\phi_{m+1}$ and $\phi_{m-1}$ is eaten by the $W$.

\section{Higgs Data}
\label{sec:app}
\begin{table}[hc]
\begin{center}
\begin{tabular}{| l | c | c | c | c |}
\hline
Channel & ($\mu_V,\,\mu_F$) & ($\Delta\mu_V,\,\Delta\mu_F$) & $\rho$ & Reference\\\hline
ATLAS $\gamma\gamma$  & (1.75, 1.62) & (1.25, 0.63) & -0.17 & \cite{ATLAS-CONF-2013-012}\\
CMS $\gamma\gamma$  & (1.48, 0.52) & (1.33, 0.60) & -0.48 & \cite{CMS-PAS-HIG-13-001} \\\hline

ATLAS ZZ & (1.2, 1.8) & (3.9, 1.0) & -0.3 & \cite{ATLAS-CONF-2013-013}\\
CMS ZZ & (1.7, 0.8) & (3.3, 0.6) & -0.7 &\cite{Chatrchyan:2013mxa}\\\hline

ATLAS WW & (1.57, 0.79) & (1.19,  0.55) & -0.18& \cite{ATLAS-CONF-2013-034}\\
CMS WW & (0.71, 0.72) & (0.96,  0.32) & -0.23 & \cite{CMS-PAS-HIG-13-005}\\\hline

ATLAS $\tau\bar{\tau}$ & (1.50, 1.04) & (1.05,  1.83) & -0.50 & \cite{ATLAS-CONF-2013-108}\\
CMS $\tau\bar{\tau}$ & (1.55, 0.66) & (1.26,  1.21) & -0.45 & \cite{CMS-PAS-HIG-13-005}\\\hline

Combined $Vh,\,h\to b\bar{b}$ & (0.9, -) & (0.3, -) & - & \cite{ATLAS-CONF-2013-079,CMS-PAS-HIG-13-012,TEVNPH:2012ab}\\
Combined $t\bar{t}h,\,h\to b\bar{b}$ & (-, -0.1) & (-,1.8) & - & \cite{ATLAS-CONF-2012-135,CMS-PAS-HIG-13-019}\\ \hline

\end{tabular}
\caption{The signal strengths with their uncertainties and correlations  for the 126 GeV resonance used in the fit.}
\label{tab:mu}
\end{center}
\end{table}

\begin{table}[tc]
\begin{center}
\begin{tabular}{| c | c | c | c | c |}
\hline
Collaboration & Channel & $\sqrt{s}\, [\text{TeV}]$ & $\mathcal{L}\, [\text{fb}^{-1}]$ & Range $m_{h^{\prime}}$ probed [GeV] \\ \hline
{\color{red}ATLAS}~\cite{ATLAS-CONF-2013-013} & $h^{\prime} \to ZZ \to 4\ell$ & $8$ & $20.7$ & $110-1000$ \\
{\color{orange}ATLAS}~\cite{ATLAS-CONF-2013-067} & $h^{\prime} \to WW \to 2(\ell \nu)$ & $8$ & $20.7$ & $260-1000$ \\
{\color{magenta}CMS}~\cite{CMS-PAS-HIG-12-024} & $h^{\prime} \to ZZ \to 2\ell 2q$ & $7+8$ & $5.3+19.6$ & $230-600$ \\
{\color{green}CMS}~\cite{CMS-PAS-HIG-13-002} & $h^{\prime} \to ZZ \to 4\ell$ & $7+8$ & $5.1+19.6$ & $100-1000$ \\
{\color{blue}CMS}~\cite{CMS-PAS-HIG-13-003} & $h^{\prime} \to WW \to 2(\ell \nu)$ & $7+8$ & $4.9+19.5$ & $100-600$ \\
{\color{cyan}CMS}~\cite{CMS-PAS-HIG-13-008} & $h^{\prime} \to WW \to \ell \nu q q^{\prime}$ & $8$ & $19.3$ & $600-1000$ \\
{\color{purple}CMS}~\cite{CMS-PAS-HIG-13-014} & $h^{\prime} \to ZZ \to 2\ell 2\nu$ & $7+8$ & $5.0+19.6$ & $200-1000$ \\ \hline
\end{tabular}
\caption{Searches for heavy Higgs boson production in the $WW$ and $ZZ$ channels used in our analysis.}
\label{tab:comb}
\end{center}
\end{table}

In this appendix we list the Higgs data used in our analysis in Tab.~\ref{tab:mu}. For each decay mode, we extract the signal strengths in the gluon-fusion plus $t\bar th$\footnote{In our analysis, we ignore the $t\bar th$ which is negligible compared to gluon-fusion production.} ($\mu_F$) and vector-boson-fusion plus $Vh$ ($\mu_V$) production channel from the reported 2-dimensional ellipses by both the ATLAS and CMS collaborations.
This allows us to capture the correlation, $\rho=\text{cov}[\mu_V,\mu_F]/(\Delta\mu_V\Delta\mu_F)$, between the two production channels.   
However, for the $b\bar{b}$ decay mode in which the correlation is absent, we also include signal strength measured by CDF and D\O. 

\begin{figure}
  \centering
\includegraphics[width=0.5\textwidth]{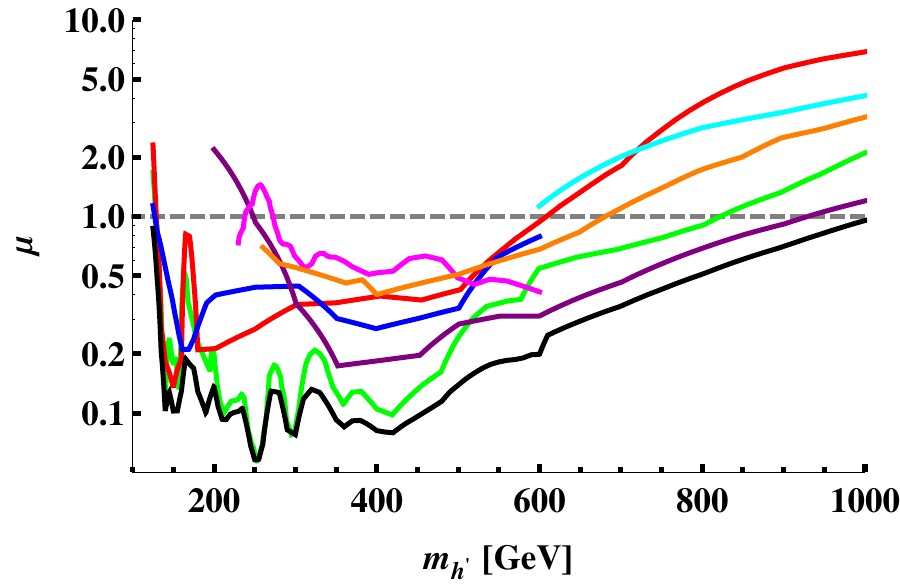}
\caption{(Colors) 95\% CL upper limits on Higgs boson signal strength in the $WW$ and $ZZ$ channels from the ATLAS and CMS collaborations. The color of the curve corresponds to the search in Tab.~\ref{tab:comb} with the same color. (Black) Our combination of those limits.}
 \label{fig:comb}
\end{figure}

In addition, we list  in Tab.~\ref{tab:comb} the searches for heavy Higgs boson production in the $WW$ and $ZZ$ channels used in our analysis. In the absence of any additional resonances, these searches set a 95\% CL upper limit on the signal strength of a Higgs boson in the $WW+ZZ$ channels. The experimentally determined signal strengths, $\mu_i$, for a given Higgs mass, $m_{h^{\prime}}$, were added in inverse quadrature, $1 / \mu^2(m_{h^{\prime}}) = \sum_i 1 / \mu^2_i(m_{h^{\prime}})$, to get the combined signal strength, $\mu$. The result of our combination is shown in Fig.~\ref{fig:comb}. We stopped the analysis at 128 GeV as that is where the expected sensitivity to additional Higgs bosons starts to degrade due to the presence of the 126 GeV Higgs~\cite{CMS-PAS-HIG-13-016}. Taking this combination at face value, a Higgs boson that is produced and subsequently decays to $WW$ and $ZZ$ with SM strength is ruled out in the interval $m_{h^{\prime}} = 128-1000$ GeV. Of course, if there is more than one Higgs, then none of them need couple to EW gauge bosons with SM strength. Instead, for a given set of parameters in a model, one must compare the predicted signal strength against the curve in Fig.~\ref{fig:comb} to determine the range(s) for which $m_{h^{\prime}}$ is ruled out.

\section{Dispersion relations}
\label{sec:appb}
Here we derive the dispersion relations used in Sec. \ref{disper-rels}. The presentation we give is close to that of
\cite{Distler:2006if}.\footnote{The presentation in the v1 of the
  arXiv version of Ref.~\cite{Distler:2006if} contains more detail
  than the published work.} The massless limit
of the massive case is reproduced by the method of
\cite{Low:2009di}. 

\begin{figure}
  \centering
 \includegraphics[width=0.55\textwidth]{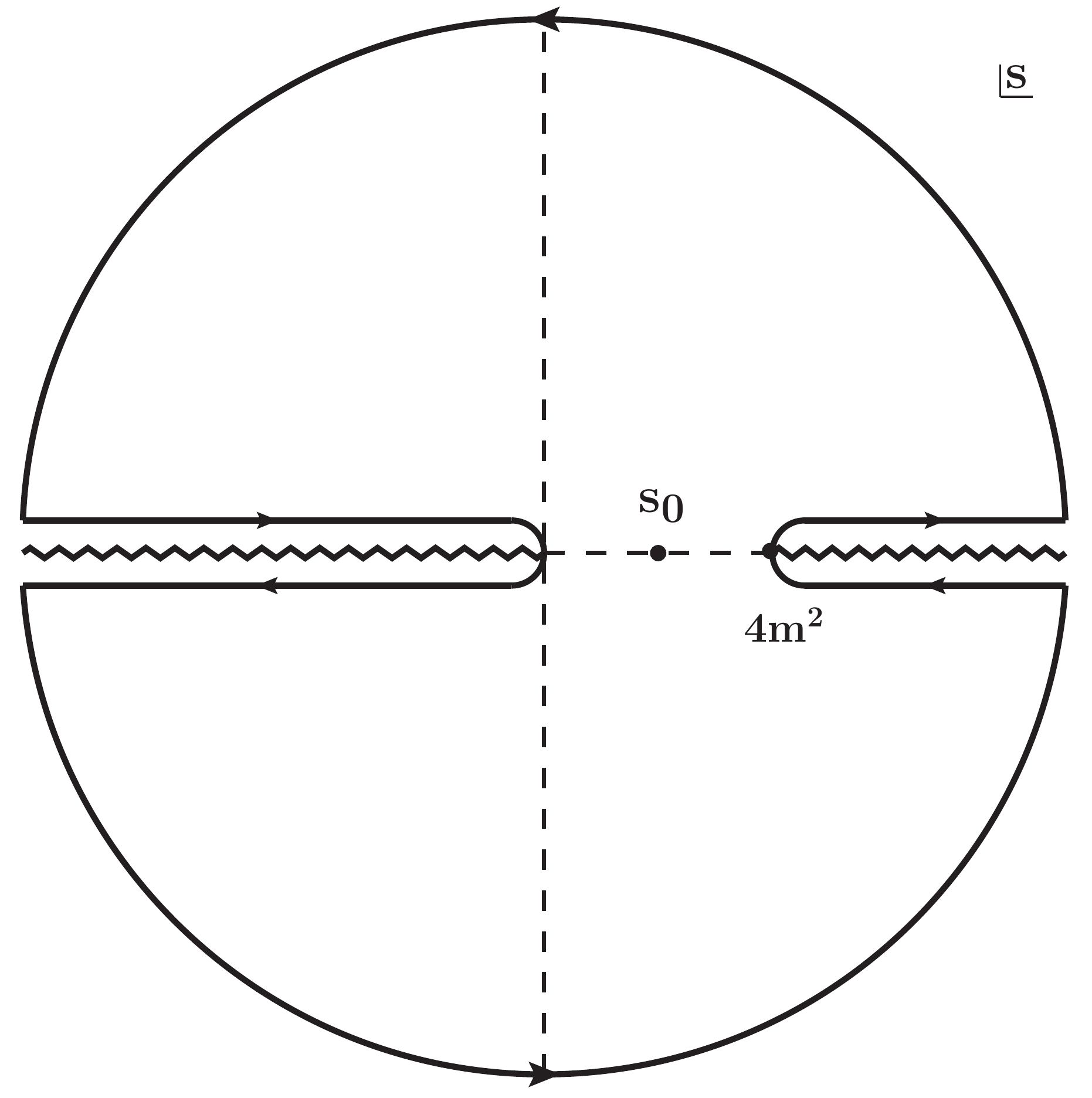}
\caption{Contour in the complex-$s$ plane for the integral that leads to the dispersion relation in equation~\eqref{eq:disp0}. The cut on the real $s$-axis runs from negative infinity to the origin and again from $4m^2$ to positive infinity. }
 \label{fig:contour}
\end{figure}

Consider a forward scattering amplitude $\CAf(s)$ for two scalar particles of a small mass $m$, to be eventually set to 0. The amplitude is only a function of the Mandelstam variable $s$ since
for forward scattering $t=0$. As a function of complex-$s$, $\CAf$ has
two branch cuts, extending along the real axis from $-\infty$ to $0$
and from $4 m^2$ to $\infty$. Cauchy's theorem applied to
this function, using the contour of Fig.~\ref{fig:contour} gives
\begin{align}
\CAf(s_0)&=\frac1{\pi}\left[\int_{-\infty}^0ds\,\frac{\text{Im}\CAf(s)}{s-s_0}
+\int_{4m^2}^\infty ds \,
\frac{\text{Im}\CAf(s)}{s-s_0}\right]\\
\label{eq:disp0}
&=\frac1{\pi}\int_{4m^2}^\infty ds \,\left[
\frac{\text{Im}\CAf(s)}{s-s_0}
+\frac{\text{Im}\CAf(4m^2-s)}{4m^2-s-s_0}\right],
\end{align}
where $2i\text{Im}\CAf(s)=\CAf(s+i0^+)-\CAf(s-i0^+)$. We have neglected
the contribution from the circle at infinity, which is justified
provided $|\CAf(s)|\to0$ as $|s|\to\infty$ faster than $1/|s|$. If
only $|\CAf(s)/s|\to0$ as $|s|\to\infty$ faster than $1/|s|$  we can
use a once subtracted dispersion relation instead, obtained from the
first derivative of the above:
\begin{equation}
\frac{d\CAf(s_0)}{ds_0}
=\frac1{\pi}\int_{4m^2}^\infty ds \,\left[
\frac{\text{Im}\CAf(s)}{(s-s_0)^2}
+\frac{\text{Im}\CAf(4m^2-s)}{(4m^2-s-s_0)^2}\right].
\end{equation}
The Froissart bound guarantees that at least $|\CAf(s)/s^2|\to0$ as
$|s|\to\infty$ faster than $1/|s|$ so a doubly subtracted dispersion relation
(obtained from one further derivative) is always possible:
\begin{equation}
\frac{d^2\CAf(s_0)}{ds_0^2}
=\frac2{\pi}\int_{4m^2}^\infty ds \,\left[
\frac{\text{Im}\CAf(s)}{(s-s_0)^3}
+\frac{\text{Im}\CAf(4m^2-s)}{(4m^2-s-s_0)^3}\right].
\end{equation}

The second term in \eqref{eq:disp0} can be related to the first using
crossing relations. Suppose the amplitude for $ab\to cd$ is
$\CA_{ab\to cd}(s,t,u)$. Then one may exchange the roles of $a$ and
$c$ by replacing their in/out momenta, $p_a\to -p_a$ and $p_c\to -
p_c$, to obtain $\CA_{cb\to ad}(s,t,u)=\CA_{ab\to
  cd}(u,t,s)$. Similarly, $\CA_{db\to ca}(s,t,u)=\CA_{ab\to
  cd}(t,s,u)$. For an application to forward scattering, we set $c=a$
and $b=d$ and $t=0$. Then $\CAf(s)=\CA_{ab\to
  ab}(s,0,4m^2-s)$, and crossing gives $\CA_{ab\to
  ab}(s,0,4m^2-s)=\CA_{ab\to ab}(4m^2-s,0,s)$ or
simply $\CAf(4m^2-s)=\CAf(s)$. Note however that for the
discontinuity across the cut we have
$\text{Im}\CAf(4m^2-s)=-\text{Im}\CAf(s)$. Using this in
\eqref{eq:disp0} one obtains
\begin{equation}
\CAf(s_0)
=\frac1{\pi}\int_{4m^2}^\infty ds \,\left[
\frac{\text{Im}\CAf(s)}{s-s_0}
-\frac{\text{Im}\CAf(s)}{4m^2-s-s_0}\right].
\end{equation}

The optical theorem may be used to relate the discontinuity across the
cut of the forward scattering amplitude to the total cross section,
$\sigma(s)$:
\begin{equation}
  \CAf(s_0)
  =\frac1{\pi}\int_{4m^2}^\infty ds \,\lambda^{1/2}(s,m^2,m^2)\sigma(s)\left[
    \frac{1}{s-s_0}
    -\frac{1}{4m^2-s-s_0}\right],
\end{equation}
where $\lambda(s,m^2,m^2)=s(s-4m^2)$. 

We are particularly interested in the case where the particles $a, b,
c$ and $d$ in the collision are identical and they carry internal
quantum numbers that correspond to elements of an irreducible
representation of a symmetry group. In fact, since we will focus on
scattering of weak interaction vector bosons, the symmetry group is
weak isospin ($SU(2)$) and the representation is a triplet.  For
identical particles we can write $\CA_{ab\to
  cd}=\CA_s\delta_{ab}\delta_{cd}+\CA_t \delta_{ac}\delta_{bd}+
\CA_u\delta_{ad}\delta_{bc}$ where $\CA_s=\CA(s,t,u)$ for some
function $\CA$, and $\CA_t=\CA(t,s,u)$ and $\CA_u=\CA(u,t,s)$. Note
that $\CA(s,t,u)$ is an amplitude {\it per se,} the one corresponding
to the case, say, $a=b\ne c=d$ ({\it e.g.,} the pion scattering
$\pi^+\pi^-\to\pi^0\pi^0$). Hence there is a dispersion relation for
$\CAf(s)=\CA(s,0,4m^2-s)$, as above. Crossing symmetry imposes some
conditions on the function $\CA$. From $\CA_{cb\to
  ad}(s,t,u)=\CA_{ab\to cd}(u,t,s)$ it follows that
$\CA_s(s,t,u)=\CA_u(u,t,s)$ and $\CA_t(s,t,u)=\CA_t(u,t,s)$. The first
is automatically satisfied, while the second gives
$\CA(t,s,u)=\CA(t,u,s)$ (The other crossing relation $\CA_{db\to
  ca}(s,t,u)=\CA_{ab\to cd}(t,s,u)$ does not lead to any further constraints). The initial state in the collision amplitude
can be prepared to have definite isospin, and if isospin is conserved
the final state will automatically have the same isospin.  Linear
combinations of $\CA_{s}$, $\CA_{t}$ and $\CA_{u}$ give the scattering
amplitude for definite isospin. In the case of interest, where the
colliding particles form an $I=1$ multiplet, the amplitudes
$T^I(s,t,u)$ for scattering in the isospin $I$ state are given by
\begin{equation}
\label{eq:TofCA}
T^0=3\CA_s+\CA_t+\CA_u,\qquad T^1=\CA_t-\CA_u\qquad\text{and}\qquad
T^2=\CA_t+\CA_u,
\end{equation}
and their inverse
\begin{equation}
\label{eq:CAofT}
\CA_s=\tfrac13(T^0-T^2),\qquad
\CA_t=\tfrac12(T^1+T^2)\qquad\text{and}\qquad 
\CA_u=\tfrac12(-T^1+T^2).
\end{equation}

We are now ready to display dispersion relations for the forward
scattering isospin amplitudes $T^I_{\text{fwd}}(s)\equiv
T^I(s,0,4m^2-s)$. The $n$-times subtracted version of \eqref{eq:disp0}
gives
\begin{equation}
\label{eq:dispI}
\frac{d^nT^I_{\text{fwd}}}{ds_0^n}
=\frac{n!}{\pi}\int_{4m^2}^\infty ds \,\left[
\frac{\text{Im}T^I_{\text{fwd}}(s)}{(s-s_0)^{n+1}}
-\frac{\text{Im}T^I_{\text{fwd}}(4m^2-s)}{(4m^2-s-s_0)^{n+1}}\right].
\end{equation}
To express this in terms of total cross sections, through the use of
the optical theorem, we use the above crossing relations,
$\CA_s(4m^2-s,0,s)=\CA_u(s,0,4m^2-s)$ and \break
$\CA_t(4m^2-s,0,s)=\CA_t(s,0,4m^2-s)$. We thus have
\begin{align}
T^0_{\text{fwd}}(4m^2-s)&=(3\CA_u+\CA_t+\CA_s)(s,0,4m^2-s),\\
T^1_{\text{fwd}}(4m^2-s)&=(\CA_t-\CA_s)(s,0,4m^2-s),\\
T^2_{\text{fwd}}(4m^2-s)&=(\CA_t+\CA_s)(s,0,4m^2-s),
\end{align}
or using \eqref{eq:CAofT},
\begin{align}
T^0_{\text{fwd}}(4m^2-s)&=\tfrac13(T^0_{\text{fwd}}-3T^1_{\text{fwd}}+5T^2_{\text{fwd}})(s),\\
T^1_{\text{fwd}}(4m^2-s)&=\tfrac16(-2T^0_{\text{fwd}}+3T^1_{\text{fwd}}+5T^2_{\text{fwd}})(s),\\
T^2_{\text{fwd}}(4m^2-s)&=\tfrac16(2T^0_{\text{fwd}}+3T^1_{\text{fwd}}+T^2_{\text{fwd}})(s),
\end{align}
we finally obtain, from \eqref{eq:dispI}, 
\begin{align}
\frac{d^nT^0_{\text{fwd}}}{ds_0^n}
&=\frac{n!}{\pi}\int_{4m^2}^\infty ds \,\lambda^{1/2}\left[
\frac{\sigma^0}{(s-s_0)^{n+1}}
-\frac{\frac13(\sigma^0-3\sigma^1+5\sigma^2)}{(4m^2-s-s_0)^{n+1}}\right],\\
\frac{d^nT^1_{\text{fwd}}}{ds_0^n}
&=\frac{n!}{\pi}\int_{4m^2}^\infty ds \,\lambda^{1/2}\left[
\frac{\sigma^1}{(s-s_0)^{n+1}}
-\frac{\frac16(-2\sigma^0+3\sigma^1+5\sigma^2)}{(4m^2-s-s_0)^{n+1}}\right],\\
\frac{d^nT^2_{\text{fwd}}}{ds_0^n}
&=\frac{n!}{\pi}\int_{4m^2}^\infty ds \,\lambda^{1/2}\left[
\frac{\sigma^2}{(s-s_0)^{n+1}}
-\frac{\frac16(2\sigma^0+3\sigma^1+\sigma^2)}{(4m^2-s-s_0)^{n+1}}\right],
\end{align}
where $\lambda=\lambda(s,m^2,m^2)=s(s-4m^2)$ and $\sigma^I$ is the
total cross section for the isospin-$I$ channel and is understood to
be a function of $s$. At $n=1$, taking the limit $s_0\to 0$ and $m\to
0$ these equations reproduce the relation given in
Ref.~\cite{Falkowski:2012vh},
\begin{equation}
\frac{dT^I_{\text{fwd}}}{ds_0}
=\frac{c_I}{6\pi}\int_{0}^\infty \frac{ds}{s} \left[
2\sigma^0+3\sigma^1-5\sigma^2\right],
\end{equation}
where $c_I=2 ,1,-1$ for $I=0,1,2$, respectively. Expanding a general amplitude in the isospin basis, $\CA=\sum_I w^I T^I $, one obtains
\beq
\frac{d\CA}{ds}\bigg|_{s=0}=\frac{2w^0+w^1-w^2}{6\pi}\int_{0}^\infty \frac{ds}{s} \left[
2\sigma^0+3\sigma^1-5\sigma^2\right]~.
\eeq
Alternatively, one can express the right hand side of the last equation in terms of the cross sections corresponding to charge eigenstates. Noticing that corresponding forward scattering amplitudes are given as
\beq
\label{eq:chargeamplitudes}
\CA_{00}=\CA_s+\CA_t+\CA_u, \quad \CA_{+0}=\CA_t, \quad \CA_{++}=\CA_t+\CA_u,\quad \CA_{+-}=\CA_s+\CA_u~,
\eeq 
and using \eqref{eq:TofCA}, we obtain
\begin{align}
&\frac{d\CA}{ds}\bigg|_{s=0}=\frac{2w^0+w^1-w^2}{\pi}\int_{0}^\infty \frac{ds}{s} \left[
\sigma_{00}+\sigma_{+0}-2\sigma_{++}\right]\nn \\&=\frac{2w^0+w^1-w^2}{\pi}\int_{0}^\infty \frac{ds}{s} \left[
2 \sigma_{+-}-\sigma_{00}-\sigma_{+0}\right]=\frac{2w^0+w^1-w^2}{\pi}\int_{0}^\infty \frac{ds}{s} \left[
\sigma_{+-}-\sigma_{++}\right]. \nn
\end{align}
The last equality reduces to \eqref{eq:disp} in the case of a single light Higgs with a non-standard coupling to the charged vector bosons $a_h$.

\begin{acknowledgments}
We thank Ryan Kelley, Ian MacNeill, and Frank W\"{u}rthwein for helpful discussions regarding the analysis in~\cite{Chatrchyan:2013fea,Chatrchyan:2012sa,Chatrchyan:2012paa}. This work has been supported in part by the U.S. Department of Energy
under grant No.~DE-SC0009919 and DOE-FG02-84-ER40153. DP is supported in part by MIUR-FIRB grant RBFR12H1MW.
\end{acknowledgments}

\bibliography{twohiggs}
\bibliographystyle{utphys}
%
%
\end{document}